\documentclass[a4paper,fleqn,usenatbib]{mnras}

\usepackage{newtxtext,newtxmath}

\usepackage[T1]{fontenc}
\usepackage{ae,aecompl}

\usepackage{graphicx}	
\usepackage{amsmath}	
\usepackage{amssymb}	

\title[Constraining churning and blurring in the Galaxy]{Constraining churning and blurring in the Milky Way using large spectroscopic surveys -- an exploratory study}

\author[Sofia Feltzing et al.]{
Sofia Feltzing,$^{1}$\thanks{E-mail: sofia@astro.lu.se (SF)}
J. Bradley Bowers,$^{1}$
Oscar Agertz$^{1}$
\\
$^{1}$Lund Observatory, Department of Astronomy and Theoretical Physics, Box 43, SE-221\,00 Lund, Sweden\\
}

\date{Submitted in original form July 2019}

\pubyear{0000}

\begin{document}
\label{firstpage}
\pagerange{\pageref{firstpage}--\pageref{lastpage}}
\maketitle

\begin{abstract}
We have investigated the possibilities to quantify how much stars move in the Milky Way stellar disk due to diffuse processes (i.e. so called blurring) and due to influences from spiral arms and the bar (i.e. so called churning). To this end we assume that it is possible to infer the formation radius of a star if we know their elemental abundances and age as well as the metallicity profile of the interstellar medium at the time of the formation of the star. Using this information, coupled with orbital information derived from \textit{Gaia} DR2 data and radial velocities from large spectroscopic surveys, we show that it is possible to isolate stellar samples such that we can start to quantify how important the role of churning is. 
We use data for red giant branch stars from APOGEE DR14, parallaxes from \textit{Gaia} and stellar ages based on the C and N elemental abundances in the stars. 
In our sample, we find that about half of the stars have experienced some sort of radial migration (based solely on their orbital properties), while 10\% of the stars likely have suffered only from churning, whilst a modest 5-7\% of stars have never experienced either churning or blurring making them ideal tracers of the original properties of the cool stellar disk. To arrive at these numbers we have imposed the requirement that the stars that are considered to be churned have highly circular orbits. If we instead  require the star has moved away from its formation position and at the same time its Galacto-centric radius at formation that lays outside of its apo- as well as peri-centre today we find that about half of the stars have undergone a combination of churning and blurring. 
Our investigation shows that it is possible to put up a framework where we can begin to quantify churning and blurring an important.
Important aspects for future work would include to investigate how selection effects should be accounted for.
\end{abstract}




\section{Introduction}
\label{sect:introdisc}

It has long been understood that a star moving in the Galactic potential will be subject to transient gravitational interactions, such as passing close to another star or a giant molecular cloud or interacting with the density enhancement caused by a spiral arm \citep[e.g.,][]{1977A&A....60..263W}. Many of these mechanisms have been studied in quite some detail and have allowed us to for example understand the structure in velocity space of the stars in the Hipparcos data as the result partially of interactions with spiral arms and the bar \citep{2000AJ....119..800D}. These interactions generally causes initially circular orbits to over time become less circular -- an effect now known as ``blurring''.

The observation that the Sun is more metal-rich than the youngest stars in the solar neighbourhood has invoked the idea that the Sun potentially came from a region interior to the current position of the sun \citep{1996A&A...314..438W}. However, no reasonable set of physical interactions have been able to explain how the Sun could have undertaken this journey. \citet{2002MNRAS.336..785S} showed that another process, ``churning'', can move a star from a more or less circular orbit to another circular orbit thus erasing all memory of the past kinematic history of the star -- this means that we can not integrate the orbits of stars backwards to figure out where they came from \citep[an example can be found in][]{2016MNRAS.457.1062M}.

Since the seminal paper by \citet{2002MNRAS.336..785S} several studies have shown that the effects of radial migration (churning and blurring combined) can be substantial \citep{2008ApJ...684L..79R,2011ApJ...737....8L}.  However, although there is no doubt that the mechanisms resulting in churning and blurring are present a proper quantification of their respective importance in the Milky Way stellar disk is still lacking. This is partially due to a lack of a large enough and reliable enough data-set but also partially due to that we are still exploring efficient ways to study these effects. \citet{2018ApJ...865...96F}
provides a model that aims to quantify the global efficiency of radial migration amongst stars in the Milky Way. Using data from APOGEE DR12 \citep{2015ApJS..219...12A} they find that with a radial orbit migration efficiency of 3.6$\pm$0.1\,kpc. In this model the sun might have  a formation radius of about 5.2\,kpc. A quite substantial distance away from its present position.

In this study we explore a possible way to quantify the fraction of stars that have migrated radially in the Galaxy. We also attempt to derive the fractions of stars that have been blurred and churned, respectively. Our method borrows ideas from 
\citet{2018MNRAS.481.1645M} and \citet{1987JApA....8..123G} and is quite simple. 
What we do is to assume that there exists a model that describes how the radial metallicity gradient in the inter-stellar medium (ISM) in the Milky Way has evolved over time. Then if we know the age and the metallicity of a star it is straightforward to derive the Galacto-centric radius at which the star formed. This then allows us to calculate how far the star has moved radially in our Galaxy from when it formed till today. Combining such information with orbital data allows to further study how stars with different kinematic properties have moved -- enabling us to put up a method to constrain both churning as well as blurring. 

In this way we are able to identify the individual stars that have migrated and can, for example study the properties of blurred stars and contrast that with the properties of the churned stars. This allows, eventually, for a fine-grained understanding of the underlying stellar populations. It also allows a study of the properties of stars that have \textit{not} moved from where they formed, providing further constraints on our understanding of how the stellar disk in the Milky Way formed.

This paper is organised as follows: Sect.\,\ref{sect:data} describes the data-set used in this exploratory study; Sect.\,\ref{sect:analysis} explains the method and describe the different sets of ISM profiles we use; in Sect.\,\ref{sect:anal} we derive upper limits on how many stars have radially migrated and how many have been churned or blurred in our sample. We also discuss limitations of the sample (selection effects) and take a look at how cosmological simulations could potentially be used to provide the evolution of the ISM; Sect.\,\ref{sect:discussion} provides a summarising discussion of our results; Sect.\,\ref{sect:conclusion} concludes the paper with a summary of our findings.

\section{Data}
\label{sect:data}

\subsection{APOGEE and \textit{Gaia} data}
\label{sect:apogee}

\begin{figure}
	\includegraphics[width=\columnwidth]{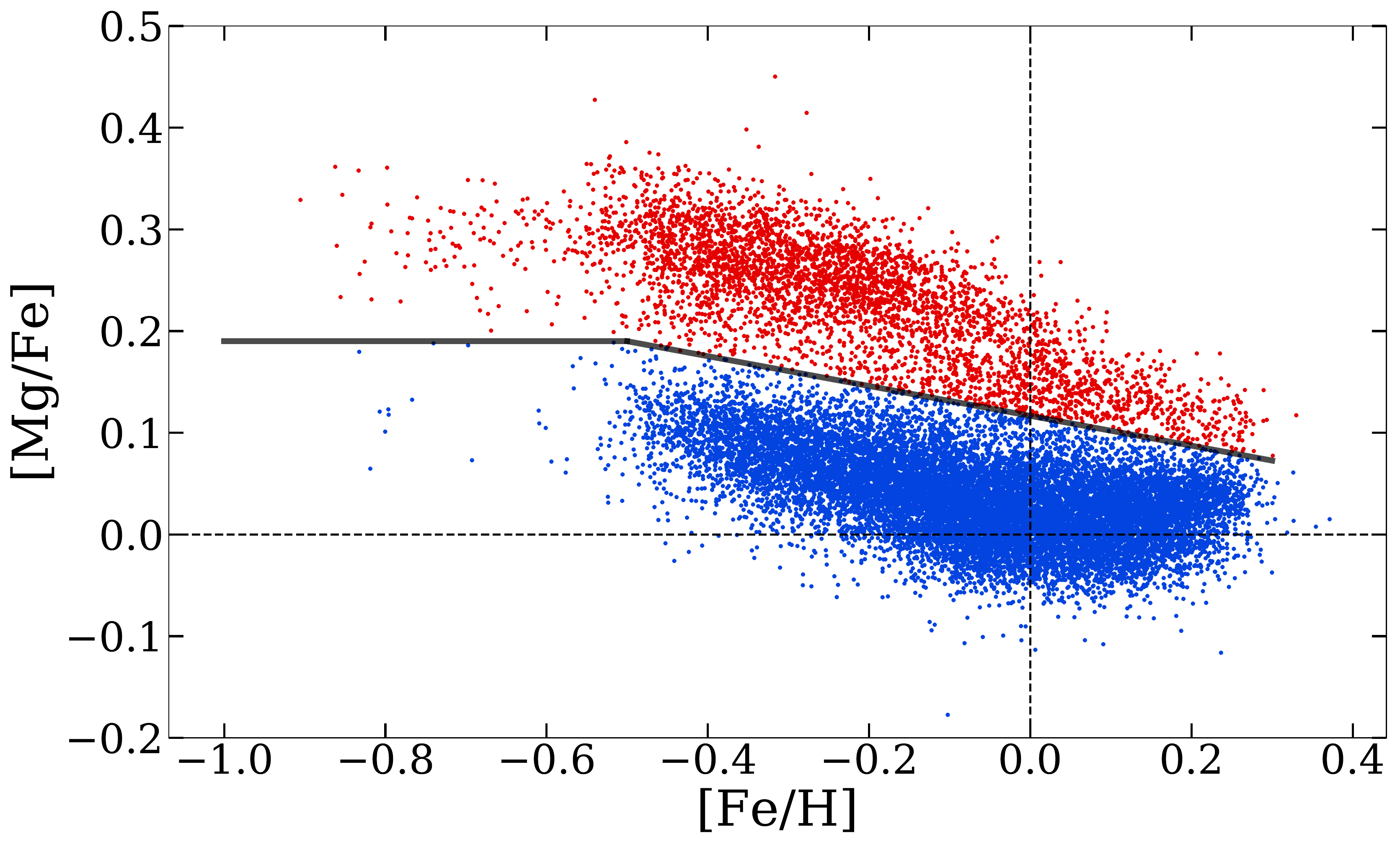}
    \caption{The APOGEE stars selected according to the four criteria listed in Sect.\,\ref{sect:apogee}. The full line shows how we divide stars in to high-$\alpha$ (blue filled circles) and low-$\alpha$ (red filled circles) stars (see Sect.\,\ref{sect:full}).}
    \label{fig:apogee}
\end{figure}

We use data from APOGEE \citep[SDSS-IV data release 14,][]{2017AJ....154...94M} and \textit{Gaia} data release 2 \citep[DR2,][]{2016A&A...595A...1G,2018A&A...616A...1G}. 
From APOGEE DR14 we select all stars that fullfil the following four criteria

\begin{enumerate}
\item They are part of the main survey targets
\item There were no failures in determining 
	\begin{itemize}
	\item the effective temperature ($T_{\rm eff}$)
	\item the surface gravity ($\log g$)
	\item the rotation
	\item the overall iron abundance ([Fe/H]\footnote{We use the standard notation where [X/H] $= \log (N_{\rm X}/N_{\rm H})_{\rm star} - \log (N_{\rm X}/N_{\rm H})_{\odot}$), X being any element.})
	\end{itemize}
\item Radial velocity could be determined for the star and the error in the radial velocity is $ < 0.5$\,km\,s$^{-1}$
\item The uncertainty in the determination of [Fe/H] is $< 0.05$\,dex
\end{enumerate}

Criterium (iii) ensures that we exclude potential binaries from the sample. 
The selected stars from APOGEE DR14 were cross-matched with \textit{Gaia} DR2. We further require that the relative uncertainty in the parallax measured by \textit{Gaia} t is less than 10\,\%. This cut allows us to do robust and straightforward calculations of the stellar orbits \citep[compare, e.g., discussion in][about selecting the best stars for studying fine-structure in the observed Herzsprung-Russel diagram]{2018A&A...616A..10G}.

The combined sample drawn from APOGEE DR14 and \textit{Gaia} DR2 fulfilling the criteria listed above amount to about 85\,000 stars. Figure\,\ref{fig:apogee} shows  [Mg/Fe] as a function of [Fe/H] for the sample. 

\subsection{Calculations of stellar orbits}
\label{sect:orbits}

Using the astrometric data from \textit{Gaia} DR2 and radial velocities from APOGEE DR14 we used {\tt galpy}\footnote{Astrophysics Source Code Library, record ascl:1411.008, https://galpy.readthedocs.io/en/v1.4.0/} \citep{2015ApJS..216...29B} to calculate velocities and orbital parameters for the stars. In parti such as the eccentricity of the orbit ($ecc.$). 

Following \citet{2012MNRAS.425.2144L} we also derive $L_{\rm z}/L_{\rm c}$ for a more robust estimate of the circularity of the orbits. Here $L_{\rm z}$ is the angular momentum in the $\hat{z}$-direction (in cylindrical coordinates) while $L_{c}$ is the angular momentum in the $\hat{z}$-direction the star would have were it on a circular orbit characterised by the same energy as the current orbit. Thus

\begin{equation}
L_{\rm c} = R_{\rm c} v_{\rm c} 
\end{equation}

\noindent
where $R_{\rm c}$ follows from solving

\begin{equation}
E = \Phi + v_{\rm c}^2 /2
\end{equation}

\noindent
where $E$ is the orbital energy and $v_{\rm c}$ is the circular velocity defined as $v_{\rm c}^2 = R\partial \Phi/ \partial R$ for $z=0$. 

The resulting distribution of $L_{\rm z}/L_{\rm c}$  are shown in Fig.\,\ref{fig:sample2} d). 
Figure\,\ref{fig:sample3} shows a comparison of $ecc.$ and $L_{\rm z}/L_{\rm c}$ for all stars in our sample. Overall the two measures correlate well, but there are deviations. Following \citet{2012MNRAS.425.2144L} we use the $L_{\rm z}/L_{\rm c}$ for the majority of our investigations, however, we apply a stricter definition of orbital circularity than they do (see Sect.\,\ref{sect:res}).

\subsection{Stellar ages}

We need stellar ages for our investigation. However, our stellar sample consists entirely of giant stars (most of them being on or near the red clump) and determining ages is difficult for such  stars. We make use of the investigation by \citet{2016MNRAS.456.3655M} who used carbon and nitrogen to infer the mass of the stars and hence provide the possibility to place the star on the right track in the HR-diagram and derive its age. These ages are good to about 40\,\%, this means that we can not study ages of individual stars but the ensemble properties should be relatively robust.

\subsection{Properties of the full sample}
\label{sect:full}

\begin{figure*}
	\includegraphics[width=\columnwidth]{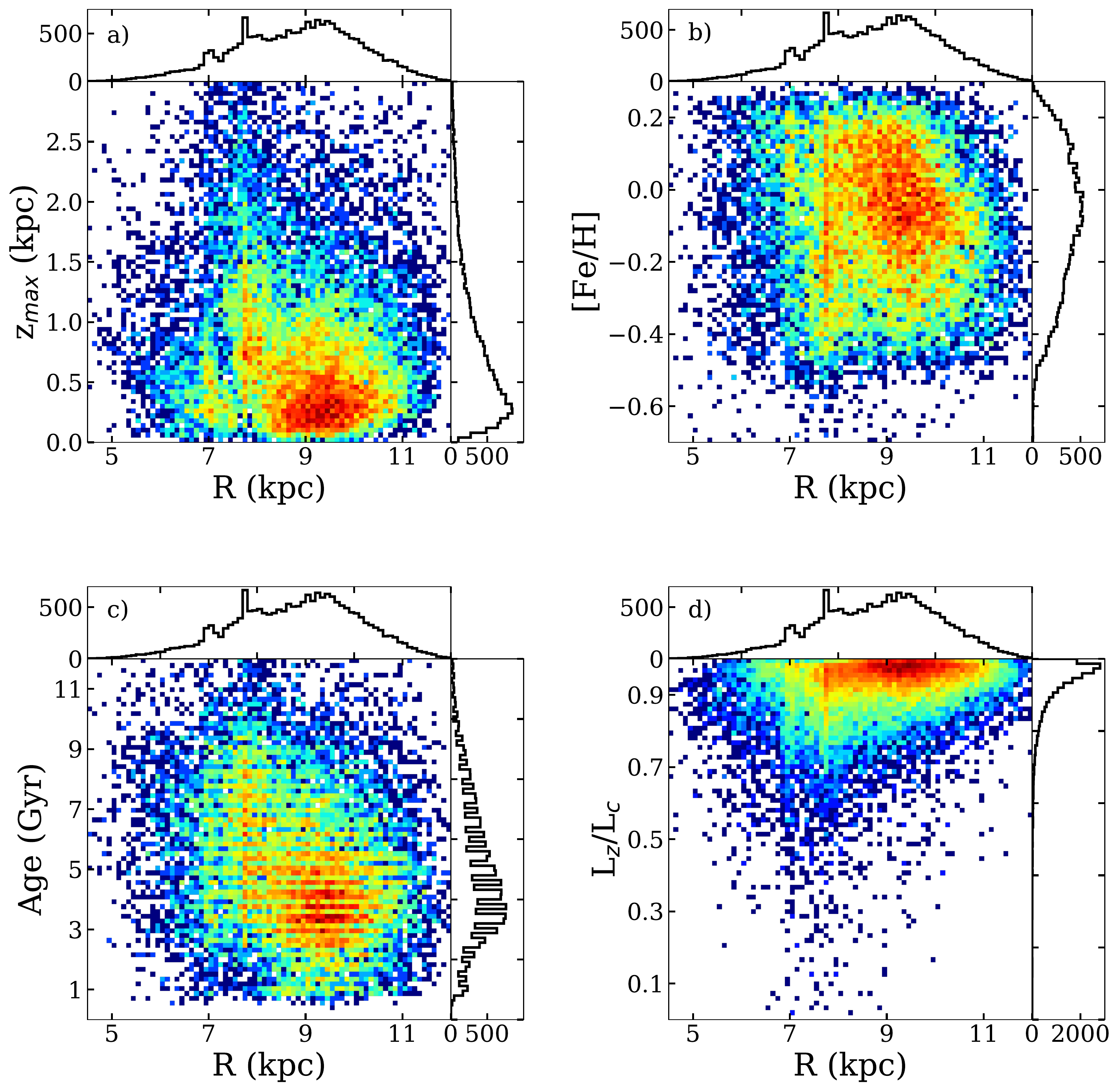}
	\includegraphics[width=\columnwidth]{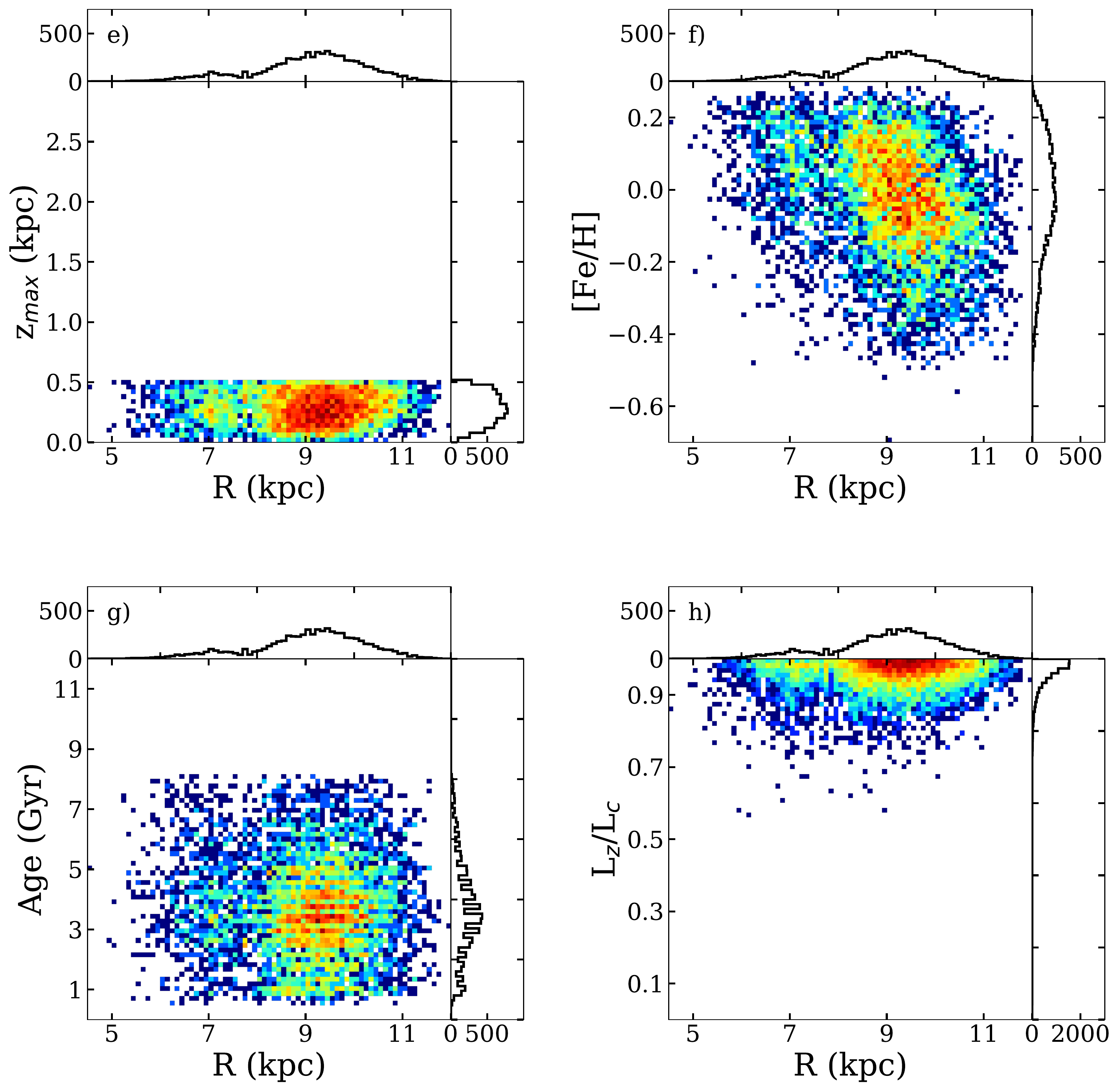}
    \caption{{\sl Left-hand panels} Characteristics of the full sample. {\bf a)} $z_{\rm max}$ as a function of current Galacto-centric distance ($R$), {\bf b)} [Fe/H]as a function of current Galacto-centric distance, {\bf c)} Age as a function of as a function of current Galacto-centric distance, {\bf d)} $L_{\rm z}/L_{\rm c}$ as a function of current Galacto-centric distance. The colour-bars show the density. {\sl Right-hand panels} Characteristics of the final sample restricted to stars in the low-$\alpha$ sequence when the cut of $z_{\rm max} < 500$\,pc has been imposed. {\bf e)} $z_{\rm max}$ as a function of current Galacto-centric distance ($R$), {\bf f)} [Fe/H] as a function of current Galacto-centric distance, {\bf g)} Age as a function of as a function of current Galacto-centric distance, {\bf h)} $L_{\rm z}/L_{\rm c}$ as a function of current Galact-ocentric distance. The colour-bars show the density of stars, with red being the highest density and blue the lowest.}
    \label{fig:sample2}
\end{figure*}

\begin{figure*}
	\includegraphics[width=12cm]{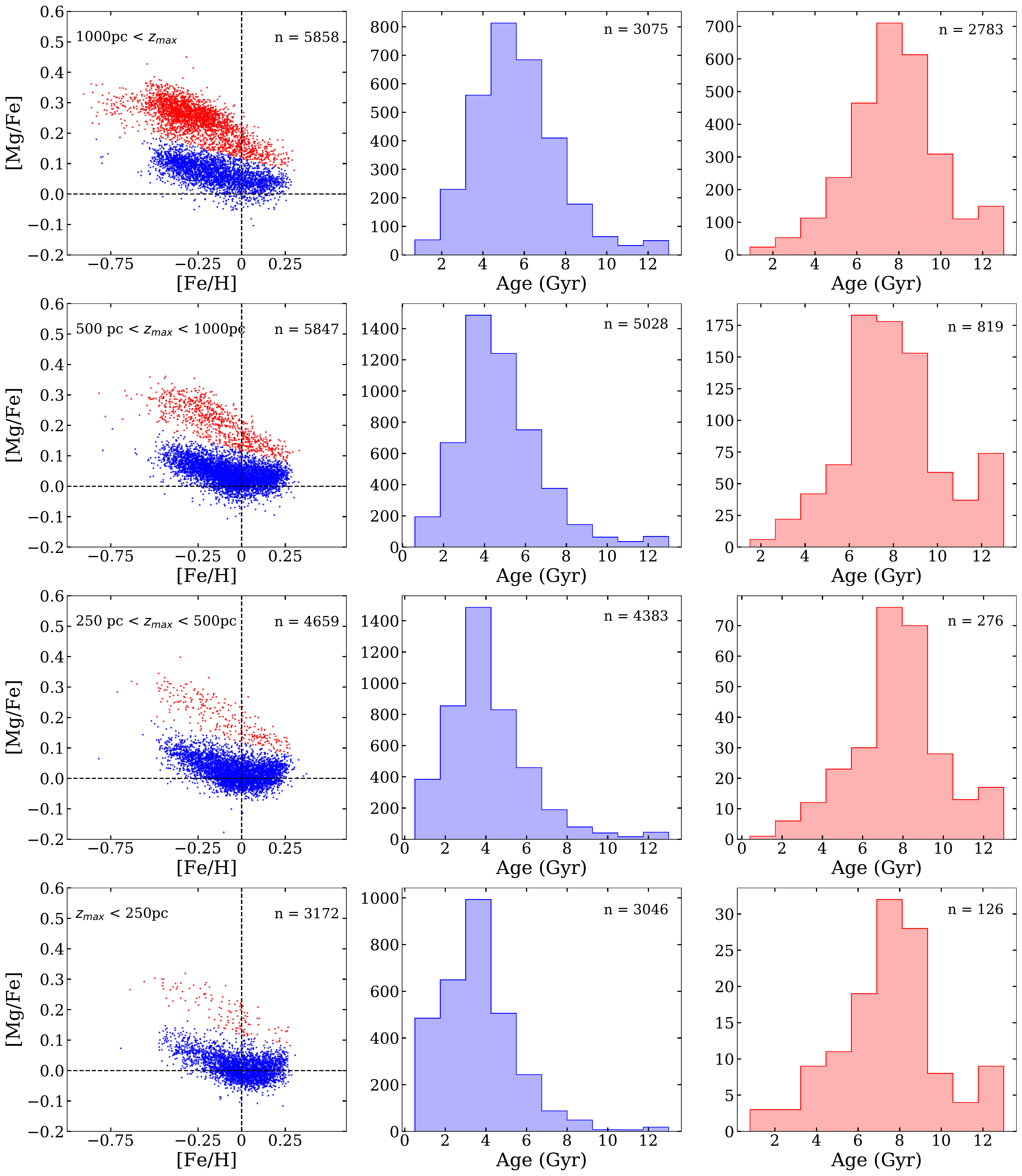}
    \caption{Stellar properties as a function of $z_{\rm max}$. To the left we show the [Mg/Fe] as a function of [Fe/H]. Colour coding as defined in Fig.\,\ref{fig:apogee}. The middle and right-hand column shows the age distributions of the the high- and low-[Mg/Fe], respectively. The sample is divided into four bins in $z_{\rm max}$: from $z_{\rm max}<250$\,pc in the bottom row to $z_{\rm max}>1000$\,pc in the top row. The number of stars in each sub-sample is indicated in the upper right-hand corner in each panel with age-histograms.}
    \label{fig:sample}
\end{figure*}

The final sample drawn from APOGEE DR14, \textit{Gaia} DR2 \citep{2018A&A...616A...1G}, and \citet{2016MNRAS.456.3655M} with stellar orbits calculated using {\tt galpy} comprises some 18\,000 stars. In this section we describe the overall characteristics of the sample as well as validate that the ages provide a reasonable description of the stars in the stellar disk. 

Figure\,\ref{fig:sample2}  show the major characteristics of our sample. Figures\,\ref{fig:sample2} a) to d) show how the full sample  covers Galacto-centric distances from about 6 to 11\,kpc and how their Galacto-centric distances relate to their other properties. 
The stars reach maximum distances above the plane ($z_{\rm max}$) of several kpc, but the majority do not reach beyond 1.5\,kpc.
Metallicities range between --0.5\,dex to super-solar and their distribution in Galacto-centric distances does not depend on metallicity.  Stellar ages ranges from 0.5 to about 10\,Gyr. The over-density of stars beyond the solar orbit have ages in the younger range whilst stars inside the solar orbit have somewhat larger ages. We will come back to these observations.

Figure\,\ref{fig:sample} provides a simple validation of the stellar ages. 
The stars have been divided in to high- and low-$\alpha$ stars, as indicated in Fig.\,\ref{fig:apogee}. Figure\,\ref{fig:sample} then shows the  sample divided into the high- and low-$\alpha$ stars  for four different ranges of $z_{\rm max}$. For each sample we also show the distributions of stellar ages for the high- and low-$\alpha$ stars.
We find that the sample of high-[Mg/Fe] stars on average are older than the low-[Mg/Fe] stars for the sub-sample closest to the plane (plots in the top row). This is expected from solar neighbourhood studies of dwarf stars \citep[e.g.,][]{2014A&A...562A..71B, 2017MNRAS.464.2610F} and thus confirms that our sample in all likelihood gives a reasonable description of the properties of Milky Way disk stars. The overall age of the low-$\alpha$ stars changes as a function of height above the Galactic plane such that the  stars  get older as we move to higher heights. The high-$\alpha$ stars, in contrast, has roughly the same median age and similar age spread at all $z_{\rm max}$. Thus resulting in an overall age-gradient for the whole sample as we move away from the plane. This is an expected behaviour.

These simple considerations serve as a validation that our ages give a reasonable representation of the stellar populations and we can use them with confidence in our investigation. 

We note that that although the high-$\alpha$ stars are sub-dominant at low $z_{\rm max}$ as expect, the relative number of low-$\alpha$ stars remains high also at large $z_{\rm max}$. At 1\,kpc there are still more low-$\alpha$ stars than high-$\alpha$ stars. The difference is not very large and could be due to selection effects. 

It is important to note that the method we use to infer formation radii for stars should only be applied to stars that are likely to have formed in a disc-like structure. It is very likely that so called high-$\alpha$ or thick disk stars have formed at a time when the interstellar medium was highly turbulent and/or in-homogenous \citep[e.g.,][]{2009ApJ...707L...1B}. 
Only later did the interstellar medium settle enough to allow for disc formation \citep{2012ApJ...758..106K}  where it is meaningful to study the effects of churning and blurring \citep[compare][]{2018ApJ...865...96F}. Figure\,\ref{fig:sample2} e) to h) show the properties of the sample when we restrict it to stars in the low-$\alpha$ trend as defined in Fig.\,\ref{fig:apogee} and with $z_{\rm max} < 0.5$\,kpc.

\section{Analysis}
\label{sect:analysis}

Formation radii were derived in the same way as in \citet{2018MNRAS.481.1645M} by assuming a model that describe how the radial metallicity gradient of the ISM changes over time and then simply referring each star to the relevant Galacto-centric distance given its [Fe/H] and age. The difference between their work and ours is that we do not aim to derive the evolution of the ISM but assume that it is known. Figure 3 in \citet{2018MNRAS.481.1645M} shows how Galacto-centric distances are assigned to stars based on their [Fe/H] and age. 

Here we will only use the stars in the low-$\alpha$ sequence as defined in as defined in Fig.\,\ref{fig:apogee} (see discussion in Sect.\,\ref{sect:full}).

\subsection{Choice of ISM profiles}

\begin{figure}
\begin{center}
	\includegraphics[width=0.8\columnwidth]{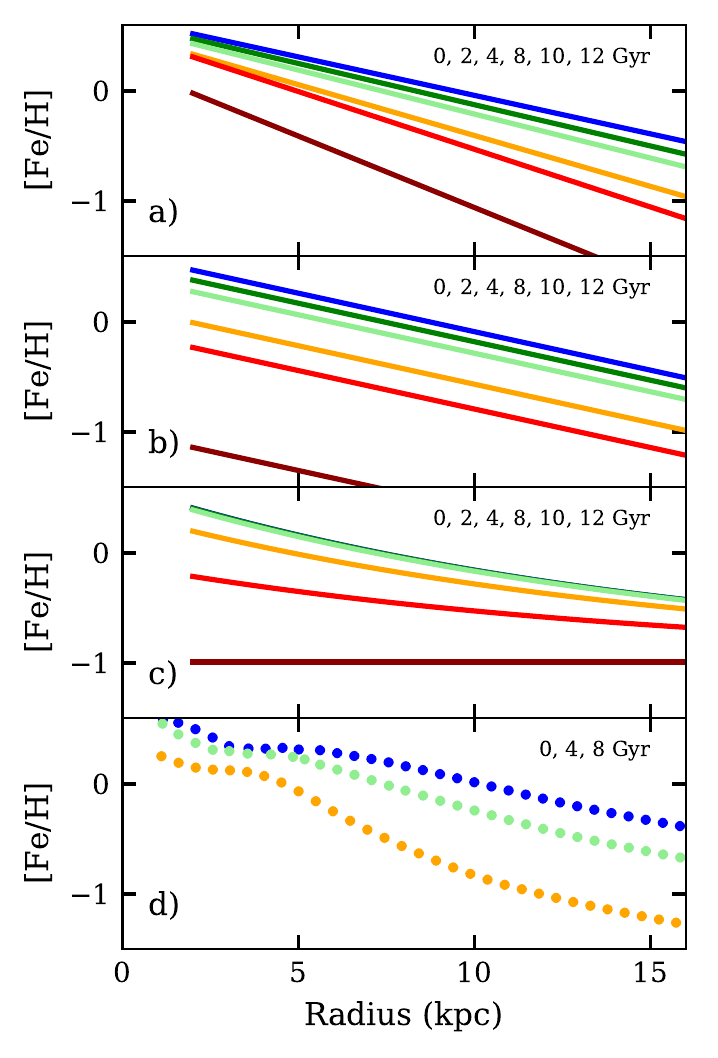}
    \caption{Comparison of radial metallicity gradients for the ISM used in this work. Ages as indicated in the legends (for coloured lines going from blue to brown with increasing age). {\bf a)} \citep{2018MNRAS.481.1645M}. {\bf b)} \citet{2018ApJ...865...96F}. {\bf c)} \citet{2015MNRAS.449.3479S}. Note that in this model the gradients for ages 0, 2, and 4 Gyr essentially overlap. {\bf d)} \citet{2015AA...580A.126K}. }
    \end{center}
    \label{fig:comparison}
\end{figure}

For our analysis we need a prescription of how the radial metallicity profile of the ISM varies over time. It is not our intention here to derive our own profiles, nor to test various profiles against observables (e.g., open cluster, O and B stars) but instead we are focusing on the method and exploring its strengths and weaknesses. Nevertheless, it is valuable to make use of a range of radial profiles to explore the method we develop. 
In the literature it is possible to find a number of radial metallicity profiles that describes how the ISM evolves over time. 

 We have  selected four sets of radial profiles that describe how the ISM evolves. The profiles are shown in Fig.\,\ref{fig:comparison}. Below we briefly summarise the major characteristics of each set, but we refer the reader to the original publications for further details \citep{2018MNRAS.481.1645M,2018ApJ...865...96F,2015MNRAS.449.3479S,2015AA...580A.126K,2015A&A...580A.127K}. We have also obtained radial metallicity profiles from an on-going cosmological simulation (Agertz et al. \textit{in prep.}). That analysis can be found in Sect.\,\ref{sect:cosm}.

\subsubsection{\citet{2018MNRAS.481.1645M}} 

As described in their paper \citet{2018MNRAS.481.1645M} derived the ISM profiles by forcing a (small) solar neighbourhood sample of stars to reproduce the distributions of formation radii for stars with different ages from the models by \citet{2013A&A...558A...9M}.  Characteristic for these radial gradients  is the steepening slope of the lines for older ages and the relatively narrow range of [Fe/H] at the smallest $R_{\rm gal}$. The radial gradients implemented in our study are shown in Fig.\ref{fig:comparison}{ a)}. 

\subsubsection{\citet{2018ApJ...865...96F} } 

\citet{2018ApJ...865...96F} present a model that parametrises the star formation history, the chemical enrichment history and profiles that show where in the Galaxy a star forms given its age and metallicity. They assume that it is possible to describe the metallicity profile of the star forming gas as a product of a radial profile and a term describing how the chemical enrichment evolves over time.  They then combine these with a model for stellar migration modelled as diffusive processes. This results in a model that can quantify the global efficiency of radial migration.

For our purposes we essentially invert their prescription of where a star forms given its age and metallicity (see Appendix\,\ref{app:iso2}). The result is radial gradients that define the evolution of the ISM with time suitable for our purposes. They are shown in Fig.\ref{fig:comparison}~b).   The gradients are straight lines that spread out more and more for the oldest ages. This is similar to one of the rejected tests by \citet{2018MNRAS.481.1645M}. 

\subsubsection{\citet{2015MNRAS.449.3479S}} 

\citet{2015MNRAS.449.3479S} set up a model of the Galaxy based on analytic distribution functions. Of interest to us is their function that describes the relation between age and metallicity for each radius in the Galaxy. The relevant function is derived by fitting to the model results from \citet{2009MNRAS.396..203S}, which includes full chemical evolution as well as gas accretion and outflow. 

Here we invert their prescription to obtain a description of how the radial gradient of the ISM evolves with time (see Appendix\,\ref{app:iso3}).
The result is shown in Fig.\ref{fig:comparison}~c).   
We  note that for young ages there is hardly any differentiation at all -- all gas share the same radial distribution at all times up and until 4\,Gyr and for the oldest age (12 Gyr) the relation is flat at --1\,dex since the model assumes that this is the metallicity of the ISM at the formation of the Galaxy \citep[see Table\,3 in][]{2015MNRAS.449.3479S}. 

\subsubsection{\citet{2015AA...580A.126K}}

 \citet{2015AA...580A.126K} built a model to study the effects of radial migration on the chemical evolution of the Milky Way. Their model includes atomic and molecular gas, star formation that depends on the molecular gas and updated SN\,Ia rates and yields. They use parametrised time-and radius-dependent diffusion coefficients to describe the radial migration. The parameterisation is based on $N$-body + SPH simulations. For further details see their paper.

Although they do not provide a tabulated description of how the metallicity gradient in the ISM evolves with time, their Fig.\,4 (bottom middle panel) show their results for three different ages. We reproduce these lines in Fig.\ref{fig:comparison}{~d)} and use them for our study. Because this model explicitly takes into account the results from the the $N$-body + SPH simulations the gradients have less idealised shapes. We use the three available lines and interpolate between them to find the formation radii of the stars. This is a simplification but one we deem reasonable as the lines appear relatively well-behaved.  

\subsection{Migratory distance -- definition}
\label{sect:md}

In order to be able to quantitatively compare and analyse the results from the different model descriptions of how the radial metallicity gradient in the ISM has varied we define the concept of migratory distance. This measure is simply the difference between the star's current Galacto-centric distance and the Galacto-centric distance at its formation:

\begin{equation}
MD = (R_{\rm current} - R_{\rm formation}).
\end{equation}

\subsection{Error on inferred formation radii}
\label{sect:error}

The principle to infer the formation radius of a star is simple, but we also need to consider the uncertainties in the properties used to derive the formation radius. In particular, we need to consider uncertainties in [Fe/H] and age as well as in stellar coordinates, parallax, proper motions and radial velocities, all of which transfer into the error of the current position of the star in the Galaxy.

For errors on [Fe/H] we use the individual errors as reported in APGOEE DR14 and for ages the error is 40\% according to \citet{2016MNRAS.456.3655M}. For the Galacto-centric radii we find that an error of 5\% being the maximum of our derived distances taking parallax errors into account. For most stars the error is between 1 and 3\%.

We include these errors in our analysis in the following way. Iterating through each star in the full data set, we create a new star by assuming a normal error distribution and drawing new parameters for age, Galactic radius, and metallicity from a Gaussian where the mean is the original observed value and the standard deviation the associated uncertainty for each variable. We do this until we have generated 10\,000 variants of each star and therefore 10\,000 new and unique data sets.

We then proceed to analyse the data just as we would have done for the original sample but now with a sample where statistical uncertainties can be readily estimated. 

\subsection{Results}
\label{sect:res}

\begin{table*}
	\caption{Fractions of stars on different types of orbits for different age bins and for 
	all stars (last column).  Results for a cut at 0.95 in $L_{\rm z}/L_{\rm c}$ can be found in Appendix\,\ref{app:95}. See Sect.\,\ref{sect:churn} for a description and Fig.\,\ref{fig:churn} for a compilation of the results.}
	\label{tab:result}
	\begin{tabular}{lccccc} 
		\hline
	 & \multicolumn{5}{l}{Fraction of stars that have $L_{\rm z}/L_{\rm c} > 0.99$ }\\
	Model	& Age $<2$ & $2<$ Age $<4$ & $4<$ Age $<6$ & $6<$ Age $<8$ & All ages\\
		\hline
		\citet{2018MNRAS.481.1645M} & 0.178$\pm$0.007 & 0.147$\pm$0.006 & 0.129$\pm$0.008 & 0.112$\pm$0.013 & 0.147$\pm$0.002\\
		\citet{2018ApJ...865...96F} & 0.186$\pm$0.006 & 0.156$\pm$0.006 & 0.141$\pm$0.009 & 0.134$\pm$0.017 & 0.160$\pm$0.002\\
		\citet{2015MNRAS.449.3479S} & 0.191$\pm$0.006 & 0.154$\pm$0.006 & 0.128$\pm$0.009 & 0.108$\pm$0.015 & 0.155$\pm$0.001\\
		\citet{2015AA...580A.126K}  & 0.170$\pm$0.007 & 0.146$\pm$0.006 & 0.130$\pm$0.007 & 0.127$\pm$0.014 & 0.145$\pm$0.002\\
		\hline
	 & \multicolumn{5}{l}{Fraction of stars that have $L_{\rm z}/L_{\rm c} > 0.99$  and outside present day orbit}\\
	Model	& Age $<2$ & $2<$ Age $<4$ & $4<$ Age $<6$ & $6<$ Age $<8$ & All ages\\
		\hline
		\citet{2018MNRAS.481.1645M} & 0.097$\pm$0.006 & 0.104$\pm$0.005 & 0.101$\pm$0.007 & 0.088$\pm$0.011 & 0.101$\pm$0.002\\
		\citet{2018ApJ...865...96F} & 0.126$\pm$0.006 & 0.123$\pm$0.006 & 0.116$\pm$0.008 & 0.112$\pm$0.016 & 0.122$\pm$0.022\\
		\citet{2015MNRAS.449.3479S} & 0.154$\pm$0.006 & 0.121$\pm$0.005 & 0.101$\pm$0.008 & 0.086$\pm$0.013 & 0.123$\pm$0.002\\
		\citet{2015AA...580A.126K}  & 0.087$\pm$0.006 & 0.097$\pm$0.005 & 0.106$\pm$0.007 & 0.115$\pm$0.014 & 0.099$\pm$0.002\\
	\hline			
	\end{tabular}
\end{table*}

\begin{figure}
   \includegraphics[width=0.95\columnwidth]{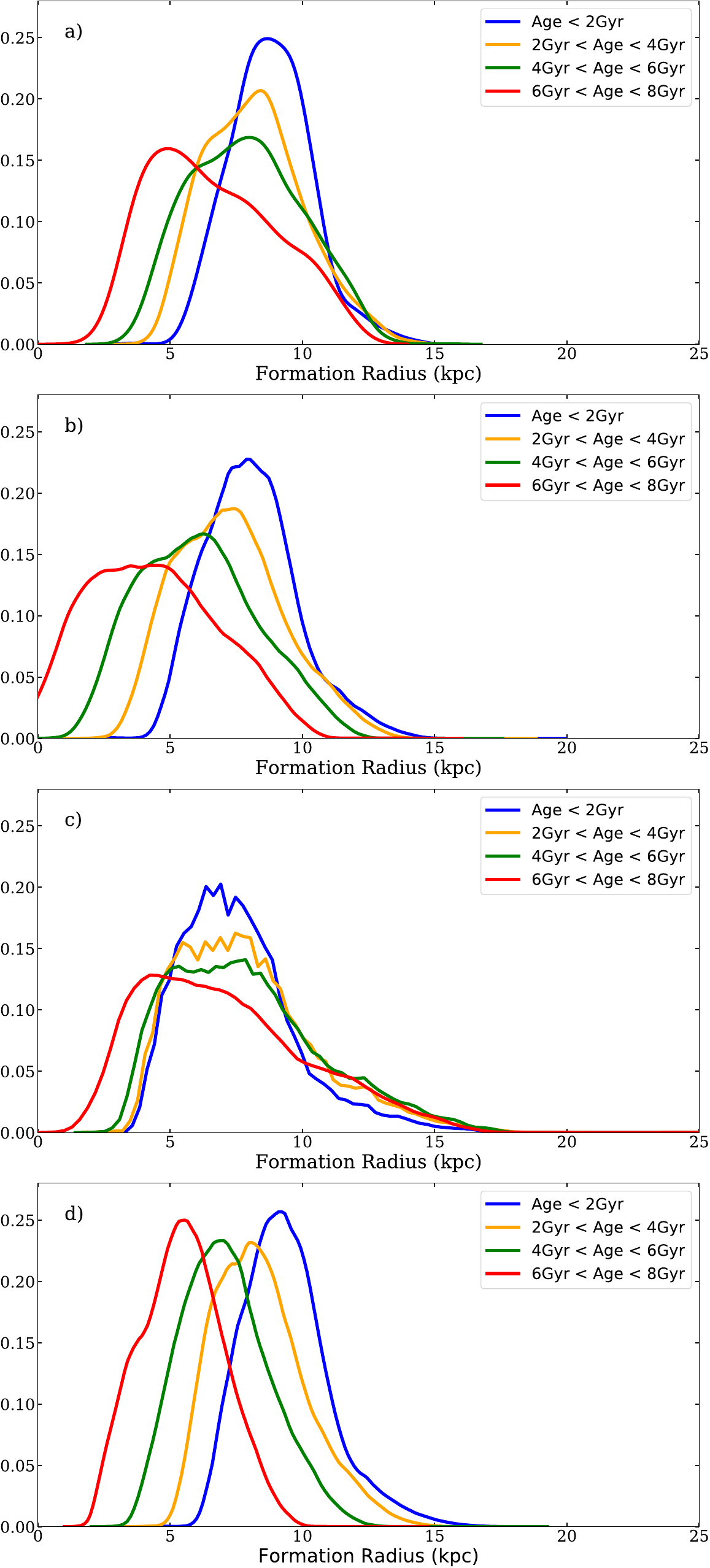}
\caption{Resulting distributions of $R_{\rm formation}$ for stars with different ages. {\bf a)} \citet{2018MNRAS.481.1645M}, ages as indicated in the legend. {\bf b)} \citet{2018ApJ...865...96F}, ages as indicated in the legend. {\bf c)} \citet{2015MNRAS.449.3479S}, ages as indicated in the legend. {\bf d)} \citet{2015AA...580A.126K}, ages as indicated in the legend. }
    \label{fig:rdistr}
\end{figure}

\begin{figure*}
	\includegraphics[width=12cm]{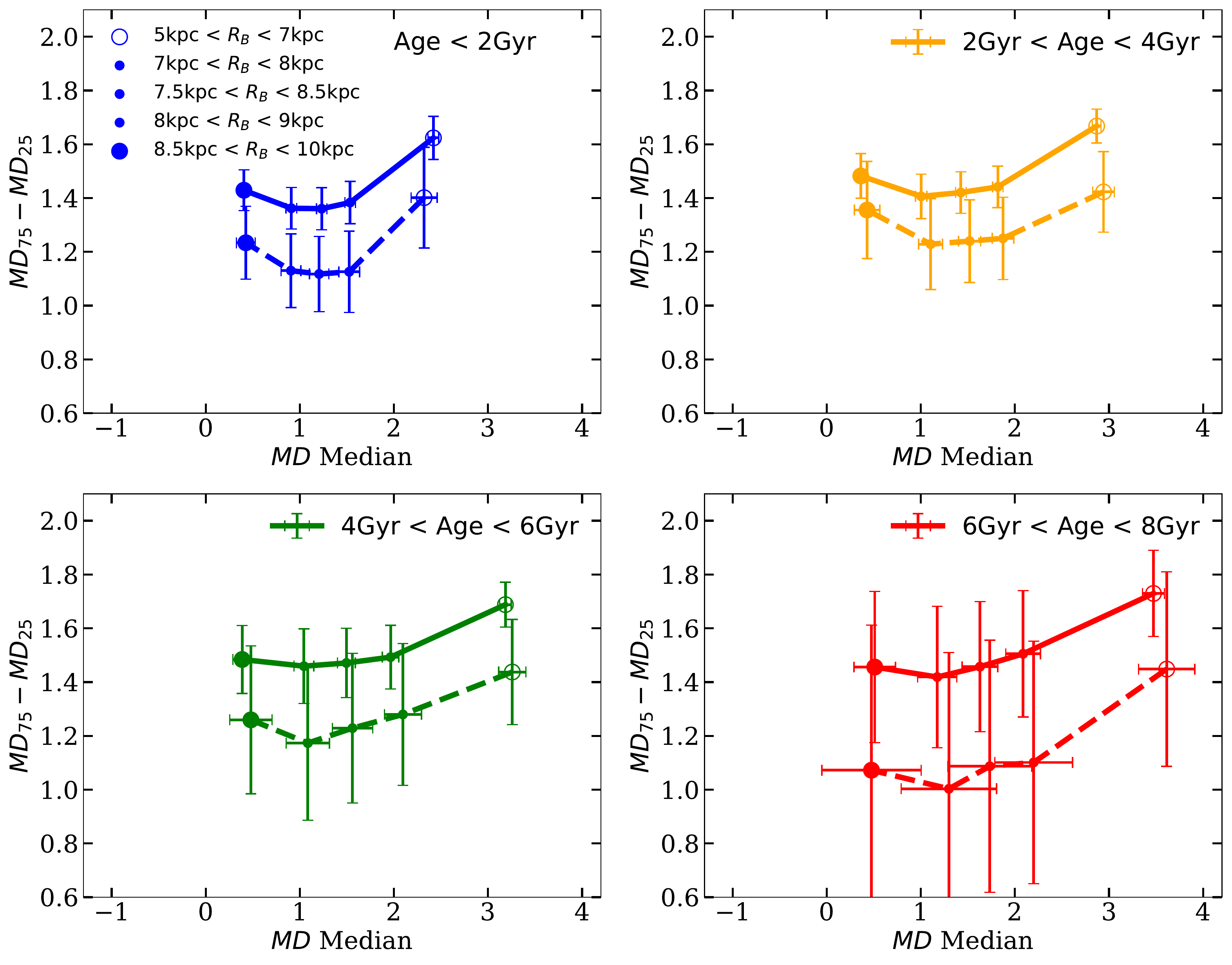}
\caption{Migratory distances ($MD$) as a function of age and formation radius (as indicated in the legend) using the model by \citet{2018ApJ...865...96F}. On the $y$-axis is shown the difference between the upper and lower quartile of the spread of the $MD$s. The solid line shows the result for the full sample, while the dashed line shows the results for stars on highly circular orbits, i.e. $L_{\rm z}/L_{\rm c} < 0.99$.}
    \label{fig:md}
\end{figure*}

Figure\,\ref{fig:rdistr} shows the resulting formation radii for stars of different ages for the four different sets of radial gradients. We note that \citet{2018MNRAS.481.1645M},  \citet{2018ApJ...865...96F}, and \citet{2015AA...580A.126K} show the same overall behaviour where older stars are on average formed further in towards the Galactic centre. This is largely in accordance with the criteria used in \citet{2018MNRAS.481.1645M}, i.e. an in-side out formation of the disk where older stars  form in the inner parts of the disc. Using the radial gradients from \citet{2015MNRAS.449.3479S} on the other hand results in that the majority of the stars in our sample forming  between roughly 5 and 8\,kpc.

Figure\,\ref{fig:md} shows the migratory distances as a function of age and formation radius of the stars. To create this figure we have used the approach to include error-estimates described in Sect.\,\ref{sect:error}. Here we show the results for \citet{2018ApJ...865...96F} but all four sets of radial gradients qualitatively show the same results. 

We also analyse a sub-sample of stars on highly circular orbits. 
After some experimentation, we find that $L_{\rm z}/L_{\rm c}> 0.99$ is the best definition of a highly circular orbit for our sample. This is a somewhat arbitrary choice but is supported by Fig.\,\ref{fig:sample3}. Figure\,\ref{fig:sample2} shows that such stars are present across the full range of Galacto-centric distances and metallicities. We believe that the choice of this cut does not significantly bias our investigation.

For all four age bins we find that stars on highly circular orbits ($L_{\rm z}/L_{\rm c} > 0.99$) have a smaller spread in migratory distance, on the order of 0.2\,kpc. The median distance that a star has migrated depends on where it has formed -- stars forming in the inner part of the disk have migrated significantly further distances than stars formed close to the sun or further out in the disk.

\section{Analysis, Interpretation, and Discussion}
\label{sect:anal}

\begin{figure}
	\includegraphics[width=\columnwidth]{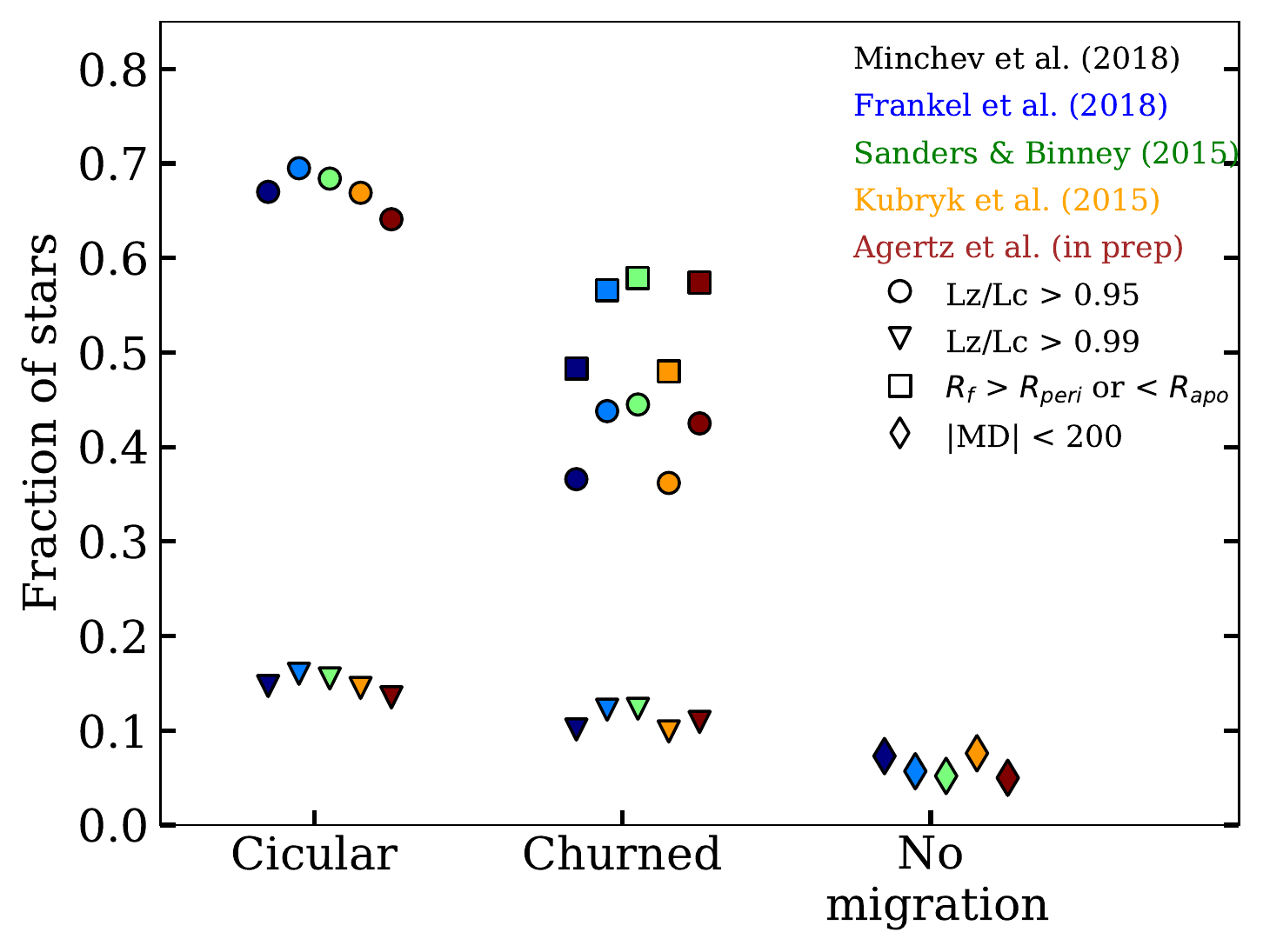}
\caption{The fraction of stars on circular orbits, of stars that have been churned and of stars that have not moved from their formation radius relative to the total sample of stars (Tables\,\ref{tab:result}, \,\ref{tab:result2} and \ref{tab:nomov}) . The colours identify the model used and the symbols different selections, see the legend. The results shown are for stars of all ages.}
    \label{fig:churn}
\end{figure}

In this section we show how combining the migratory distances with information about the stars' orbits 
can be used to constrain how many of the stars in a sample have, e.g., been churned. We discuss different ways of defining if a star has been churned or not and also look at stars that have not migrated at all. We also look at the formation radius of the sun (Sect.\,\ref{sect:sun}) and investigate the possibility to obtain the radial metallicity gradients for the ISM from a cosmological simulation (Sect.\,\ref{sect:cosm}).

\subsection{Stars that have migrated}
\label{sect:radmig}

In this section we will consider different ways to decide if a star or a stellar population has migrated in the Galaxy. We will provide some examples of how one can estimate the fractions of stars that have migrated and look at the fraction of stars that have only been churned.

Figure\,\ref{fig:churn} provides a summary of the  results.

\subsubsection{Estimating how many stars have radially migrated}

The simplest definition of a star that has radially migrated in the Galaxy is a star for which its formation radius is not the radius the star formed at. As can be inferred from Sect.\,\ref{sect:res}, the fraction of stars that have moved radially in the Galaxy is large. Adding orbital information allows for a more interesting analysis.  

Here we consider a slightly more involved criterion for stars that have migrated in the Milky Way disk; namely stars that have a Galacto-centric formation radius that is outside the Galacto-centric radial range spanned by the apo-centre and the peri-centre of the current orbit of the star:

\begin{equation}
R_{\rm formation} < R_{\rm peri, current\,orbit}\, {\rm or}\end{equation}
\begin{equation*}
R_{\rm formation} > R_{\rm apo, current\,orbit}
\end{equation*}

Since there is no requirement set on the shape of the orbit this implies that the sample of such stars must include both stars that have been only churned as well as stars that have been both churned and blurred. 
Such stars are rather common, with about half of the stars in our sample fulfilling this criterium. The results are shown in Fig.\,\ref{fig:churn}, where we also show the results from the sections below. For the models by \citet{2018MNRAS.481.1645M} and \citet{2015AA...580A.126K} we obtain a lower fraction of radially migrated stars (about 0.5) while for the other three models \citep[][and, Agertz et al. in prep]{2018ApJ...865...96F,2015MNRAS.449.3479S} we obtain a higher fraction. 

\subsubsection{Estimating how many stars have been churned}
\label{sect:churn}

In this section we explore ways to estimate how many of our stars that have experienced churning \citep[churning being the radial migration that causes a star to move radially without loosing the circularity of its orbit][see also Sect.\,\ref{sect:discussion}]{2002MNRAS.336..785S}. For this we need to define the sub-sample of stars that we think should have been subjected to just churning and not blurring. A star that has been blurred can also have been churned. This means that what we are trying to do here is to find  a conservative lowest fraction of stars that have just been churned. We define a star to be a candidate as a churned star if it has a relatively circular orbit. We define such stars as those with $L_{\rm z}/L_{\rm c} > 0.99$. 

We consider $MD$ (migratory distance) as a function of stellar age and formation radius, compare Fig.\,\ref{fig:md}.  Table\,\ref{tab:result} summaries our results. The top part of the table lists first the fraction of stars with highly circular orbits defined as $L_{\rm z}/L_{\rm c} > 0.99$ (results for stars with $L_{\rm z}/L_{\rm c} > 0.95$ can be found in Appendix,\,\ref{app:95}). We find that younger stars have a larger fraction of stars on circular orbits. This fraction decreases from about 0.18 to 0.12 as the stellar age increases. An average over all age bins gives a fraction of about 0.15.
 
Although this is an interesting number we do not think that this gives a minimum fraction of stars on churned orbits. To be churned we further require that the the star's formation radius lies outside of the range of orbits their present-day orbit occupies. As expected this selection gives a smaller fraction of stars. Again, the fraction decreases with increasing age. The total fraction of stars that these criteria and hence are prime candidates for having been churned is about 0.1 (see lower part of Table\,\ref{tab:result}). 

The results are shown in Fig.\,\ref{fig:churn} where we also include the same estimates but with a more relaxed criterion on circularity (0.95 instead of 0.99). With the more relaxed criterion on circularity for the orbits the minimum fraction of churned orbits increase to about 0.4. However, for reasons discussed below we do not believe that this gives an indication of the fraction of churned stars.

\subsubsection{Estimating the size of churning and blurring}
\label{sect:size}

It is also interesting to attempt to estimate the size of churning and blurring. An analysis as the one provided in Fig.\,\ref{fig:md} allows us to do this. It is readily seen from this figure that the spread in $MD$ for all stars is about 0.2\,kpc larger than for stars on highly circular orbits. If we regard the stars on the highly circular orbits as essentially uninfluenced by blurring then this gap is an indication of the size of the increased orbital spread due to blurring. We note that the difference between the full sample and the sample on highly-circular orbits is the same for all formation radii for a given age. From our data we can not say if the difference between samples remains constant as age increases or not. As both churning and blurring can be modelled as diffusive processes it is possible that the difference remains the same over time. All five models investigated in this study show the same patterns.

If we instead consider the $x$-axis of the plot ($MD$) we find that stars in our sample that formed inside the solar circle have migrated on average more than those that formed outside the solar circle, with a monotonic change in $MD$ as a function of formation radius. We also note a trend with age where older stars have a larger $MD$ for the same formation radius as compared to younger stars. This is consistent with radial migration being a diffusive process. Although the youngest stars have smaller $MD$ they still show substantial radial migration, indicating that churning must be a process that acts quickly on a stellar sample. We know it can not be blurring as blurring is a slow process and because even the circular sample shows significant migratory distances.

If we combine this finding with the finding that a minimum of about a tenth of the stars have been exclusively churned it indicates that about 40\% or so of the stars in our sample have migrated either by being just blurred or blurred and churned.

\subsection{Stars that have not moved}
\label{sect:nomov}

\begin{table*}
	\caption{The fraction of stars which have not moved radially in the Galaxy since they formed (see Sect.\,\ref{sect:nomov}). }
	\label{tab:nomov}
	\begin{tabular}{lccccc} 
		\hline
	& \multicolumn{4}{c}{Fraction of stars that have not moved}\\
	Model	& Age$ <2$ & $2 <$ Age $< 4$ & $4<$ Age$< 6$ & $6 <$ Age$<8$ & All ages\\
		\hline
		\citet{2018MNRAS.481.1645M} & 0.100$\pm$0.008 & 0.071$\pm$0.005 & 0.056$\pm$0.006 & 0.051$\pm$0.010 & 0.073$\pm$0.003\\
		\citet{2018ApJ...865...96F} & 0.073$\pm$0.006 & 0.054$\pm$0.005 & 0.045$\pm$0.006 & 0.039$\pm$0.011 & 0.057$\pm$0.003\\
		\citet{2015MNRAS.449.3479S} & 0.046$\pm$0.005 & 0.055$\pm$0.005 & 0.057$\pm$0.007 & 0.051$\pm$0.011 & 0.052$\pm$0.003\\
		\citet{2015AA...580A.126K}  & 0.124$\pm$0.009 & 0.081$\pm$0.006 & 0.047$\pm$0.006 & 0.020$\pm$0.007 & 0.076$\pm$0.003\\
		\hline
	\end{tabular}
\end{table*}

\begin{figure}
	\includegraphics[width=\columnwidth]{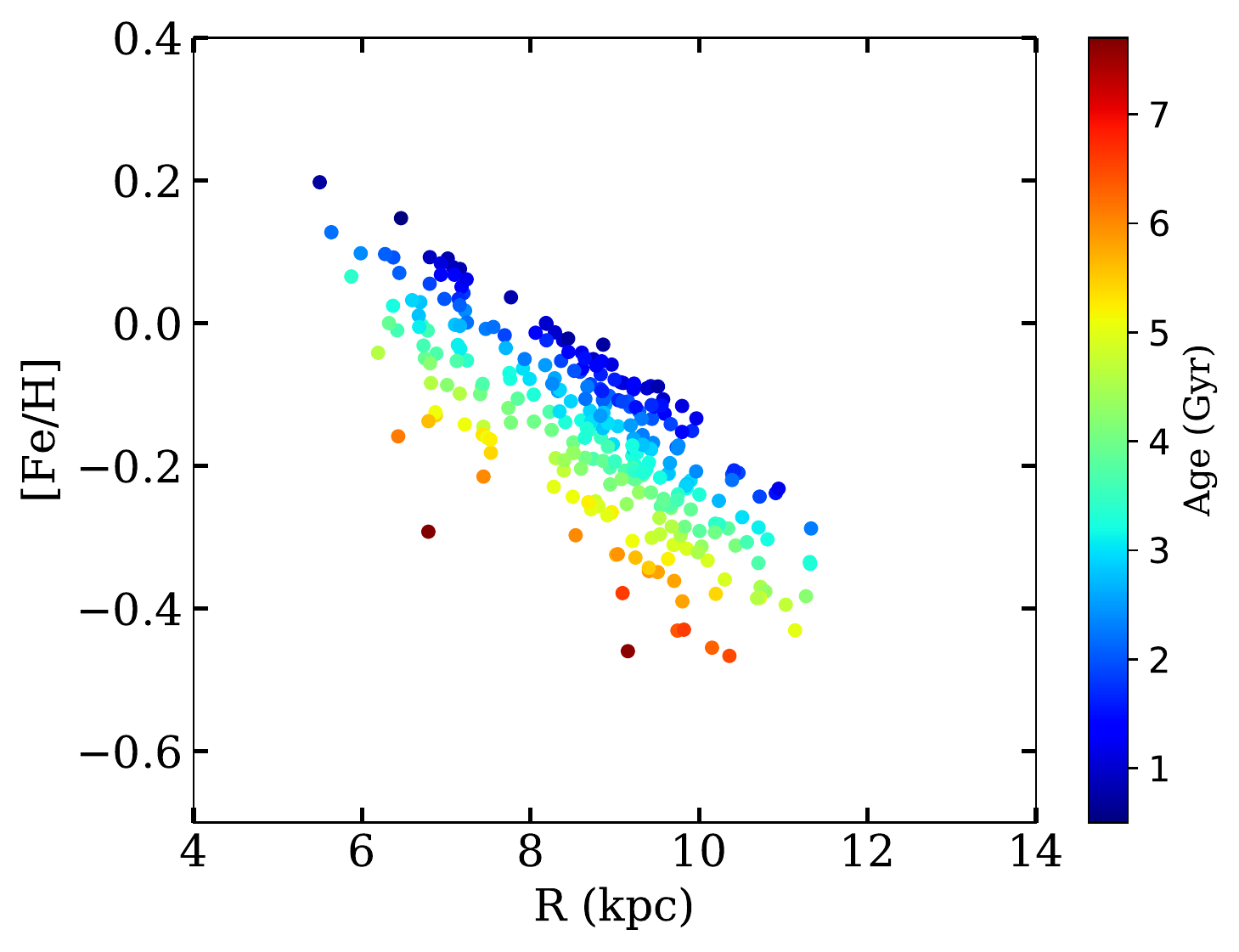}
\caption{[Fe/H] as a function of Galacto-centric radius for stars that have not moved (for definition refer to Sect.\,\ref{sect:nomov}) away from the radius they formed at. This example is for the model by \citet{2018ApJ...865...96F}. 
Age is colour coded according to the colour-bar to the right.}
    \label{fig:nomov}
\end{figure}

With our methodology it is possible to identify stars that have not moved radially in the Galaxy since they formed. We define such stars as those with $|MD| < 200$\,pc. Table\,\ref{tab:nomov} summarizes the fraction of these stars for the different models. We find that between 5 and 7\% of the stars in our sample have not moved radially in the Galaxy since they formed. The number  decreases as the age of the stars increases for all models apart from those from \citet{2015MNRAS.449.3479S} where the fraction remains constant as a function of stellar age. This is consistent with what we saw in Sect.\,\ref{sect:res} and Fig.\,\ref{fig:rdistr} where we show that this model produced roughly the same amount (and spread) in radial migration for all ages apart from the oldest stars.
It appears natural that populations of older stars should have a smaller fraction of stars that have not moved radially in the Galaxy since they formed since over time a star's position in the Galaxy will be influenced by several phenomena, not the least blurring and churning. As  time goes a star become more likely to have suffered from such phenomena and hence its orbit will start making excursions away from the original circular orbit. 

Figure\,\ref{fig:nomov} shows the properties of the sample of stars that have not moved. We note that at all radii there is a large spread in metallicity as well as age. Metallicity and age appear to correlate well at each radius, such that younger ages are associated with higher metallicities. In a slightly circular argument this could be taken as proof that indeed at a given radius there is a tight age-metallicity relation. When we look at all stars in our sample this is not the case, indeed, many studies of stars in the solar neighbourhood have shown that there is an acute lack of such a relation \citep[e.g.,][]{1993A&A...275..101E,2001A&A...377..911F,2011A&A...530A.138C}. It should be noted for our method, that although per construction more metal-rich stars are younger at a given Galacto-centric radius they need not be on orbits that have not moved via churning and/or blurring. It would have been entirely possibly that there were only stars of one age or one metallicity at a given radius that were still on the same orbit they had when they formed. Thus, we believe that this does give observational support to the assumption that stars at a given radius in the Galaxy follow a tight age-metallicity relation.

\subsection{The Galacto-centric formation radius of the Sun}
\label{sect:sun}

Returning to the question wether or not the Sun has formed at the solar radius we find that using the radial metallicity gradients for the interstellar medium from \citet{2018ApJ...865...96F}, \citet{2015MNRAS.449.3479S}, and \citet{2015AA...580A.126K}  the sun formed at a distance from the Galactic centre of 5.7, 7.0, and 6.8 kpc, respectively. Clearly, there are uncertainties associated with these estimates, however, it is also clear that if we require an inside out-formation scenario in which the metallicity in the interstellar medium is enriched over time in such a fashion as to produce flatter and flatter radial gradients then the sun most likely formed between 5.5 and 7\,kpc from the Galactic centre.
This is largely consistent with other estimates of the sun's formation radius \citep{1996A&A...314..438W,2013A&A...558A...9M,2018ApJ...865...96F}. 

\citet{2019A&A...625A.105H} discusses the possibility that the sun instead of coming from inside its current solar position in the Galaxy, it in fact has migrated in from the regions outside the sun's position. They find that this is supported by the numerical experiments carried out by \citet{2016MNRAS.457.1062M}, who uses backward integration in the Galactic potential to find out where a star comes from. As discussed earlier, with the inclusion of churning such integrations looses their ability to make such predictions, i.e. once a star has been churned you can not any more find out where it came from. As we have shown not all stars have been churned, some may just be blurred, others are untouched by dynamical encounters. So in theory the sun may have an un-churned orbit and you can retrace its orbit, however, we note that the fraction of stars in our sample fulfilling this criterium is small (compare Fig.\,\ref{fig:churn}).

\subsection{Agertz et al.\,(in prep)}
\label{sect:cosm}

\begin{figure*}
	\includegraphics[width=1.5\columnwidth]{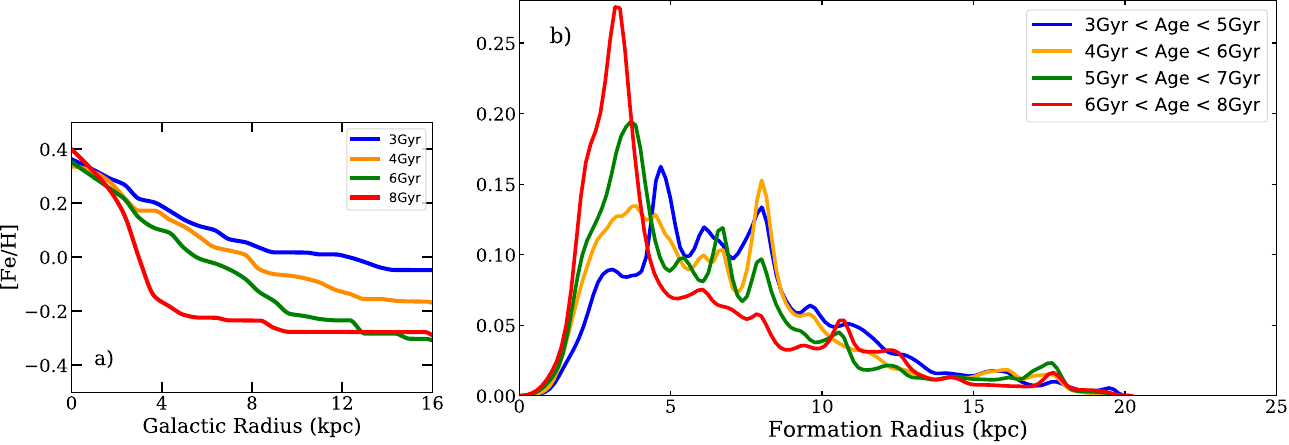}
\caption{Data based on the simulations of Agertz et al. (in prep.). a) Radial metallicity gradients in the ISM as a function of time.  b) Distribution of formation radii derived using the gradients from the simulations by Agertz et al. (in prep.). See Sect.\,\ref{sect:cosm} for further details. Note that the simulation does not (yet) go to age zero.}
    \label{fig:rdistr_cosm}
\end{figure*}

Agertz et al.\,(in prep) carried out a cosmological hydrodynamic+$N$-body zoom-in simulation of a Milky Way-mass galaxy ($M_\star\sim 6\times 10^{10}~$M$_\odot$) forming in a dark matter halo with virial mass $M_{\rm vir}=1.3\times 10^{12}~$M$_\odot$ at $z=0$ . The simulation was carried out using the adaptive mesh refinement code {\small RAMSES} \citep{teyssier02}, assuming a flat $\Lambda$-cold dark matter cosmology.  We refer to Agertz et al.\,(in prep) for details, as well as \cite{Rhodin2019} for an extensive description of the included physics and simulation settings. The simulation reaches state-of-the-art resolution, with a mean resolution of $\sim 20$\,pc in the cold interstellar medium. For every simulation snapshot we identify the most massive progenitor to the $z=0$ Milky Way-mass galaxy. We identify the disc plane from the stellar and gaseous angular momentum vector, and compute radially averaged [Fe/H]-profiles from all neutral gas within a 2\,kpc thick slab. The resulting radial ISM profiles are shown in Fig.\ref{fig:rdistr_cosm} a).

We note that insight drawn from abundance gradients in cosmological simulations of galaxy formation must be considered with care. While simulated galaxies can be selected to represent extended discs with global properties in line with the Milky Way's, their detailed assembly histories will not necessarily be compatible with that of the Milky Way galaxy. However, for this explorative study we find it informative to include these models as a part of our suite of diverse ISM models.

The radial gradients in Fig.\ref{fig:rdistr_cosm} a) share many of the overall characteristics of those derived in \citet{2018MNRAS.481.1645M} and those used in  \citet{2018ApJ...865...96F} in as much as they steepen towards the inner galaxy and older profiles have lower iron content. Thus, we can expect that overall this description of the temporal evolution of the radial metallicity profile of the ISM should give rise to similar results as found for  models explored previously. Indeed, that is also what we see, but with some modifications. Notably, Fig.\ref{fig:rdistr_cosm} b) shows that although the median radius moves to smaller and smaller radii as age increases \citep[i.e. an inside out formation scenario as imposed in][]{2018MNRAS.481.1645M} there is significant spread at all ages. 

Also for this model we have derived the fractions of stars fulfilling the different criteria discussed in Sects.\,\ref{sect:radmig} and \ref{sect:nomov}. We show these fractions in Fig.\,\ref{fig:churn} together with the results from the other models. We note that the results from this model is largely the same as for the other models.

\subsection{Discussion}
\label{sect:discussion}

One of the leading takeaways from our experimentation with this data-set is the apparently rapid onset of radial migration. We find that the median migration distances of the stars in each of our radial bins remain very constant in time. This extends out to our oldest age bin, whose stars generally show identical median migration distances to our youngest age bin. This implies that the bulk of radial migration must take place relatively early in a star’s lifetime. We suggest that this could be attributed primarily to the fact that the period of the spiral bar pattern, dominantly responsible for churning, is significantly shorter than the width of our youngest age bin. A considerable number of attempts have been made to constrain the pattern speed of the spiral arms using hydrodynamic simulations \citep[see for example][]{2007A&A...467..145C,2003MNRAS.340..949B} and observations of nearby velocity fields \citep{2001A&A...372..833F}. Most have placed the period of the spiral arms between 250 and 350\,Myr \citep{2011MSAIS..18..185G}. Thus, a spiral arm crossing is well-contained within the interval covered by our first age bin, and given that only a single transient interaction with a spiral arm is required to generate substantial changes to a stars angular momentum \citep{2002MNRAS.336..785S}, this should be sufficient to generate the displacements we observe.

This interpretation is also consistent with numerical $N$-body simulations which find substantial migration distances within 1\,Gyr due to churning. Indeed, the time-scale for churning to displace the stars can be as short as 0.5\,Gyr \citep{2016AN....337..957G}. It has also been observed that the distribution of stellar radii occupied for stars formed in the same location spreads dramatically in the first few Gyr, and slows down considerably at later times \citep{2013MNRAS.436.1479K}. Our data provide empirical confirmation of these results.

Furthermore, the swift onset of radial migration, and subsequent invariability of median motions with time, indicates that churning is the dominant mechanism through which stars initially change their Galactic radii. It has long been understood that blurring alone is not sufficient to explain the distribution of stars observed in the local region and it is now understood that that churning plays an integral part in explaining the locally observed distributions \citep[examples can be found in][]{2009MNRAS.396..203S,2013A&A...558A...9M}. Though the relative strengths of churning and blurring, that is, how much of a stars displacement from the location it formed can be described by one process or the other, has not been much studied. This is due fundamentally to the fact that these processes act simultaneously, and are largely inseparable in observation. 

It has been one of our aims in this work to investigate if it is possible to constrain the relative strengths of the two processes. Our ability to compare the migration distances of stars on circular orbits to the total sample enables us to observe how well the motions in our total sample are explained by a population that has not yet been blurred.
In doing this, we see that churning immediately has a large effect on stellar radii, while blurring accounts for a comparatively small increase in spread at this time. This result is seen in the relatively small difference in spread between our total sample and the circular subset when compared to the overall spreads of these populations. 

Blurring is traditionally modelled as a diffusive process, gradually increasing the spread of the radial distribution with time \citep{2002MNRAS.336..785S,2009MNRAS.396..203S}. As such, the deviations in radius are smaller and symmetric about the mean. In our data, this is represented by the equivalent median, though increased spread, in our total sample when compared to the  population of stars on circular orbits. The difference in spread between these populations is presumably indicative of the added diffusion from blurring. Meanwhile, the expected changes to angular momentum caused by churning are quite large -- several kpc from a single interaction with the spiral \citep{2002MNRAS.336..785S}  -- when compared to the scales of the blurring found in our data: a spread only 0.2\,kpc separating our total sample and circular subset. This leads us to believe that churning is a much stronger force than blurring in terms of the magnitude of displacement.

This is coupled with the observation that our data increases its spread with time, though only slightly from age bin to subsequent age bin. We have noted that this consistent dispersion in time is present in equal magnitude for both our total sample and our highly-circular sub-sample. This implies that the aggregate spreading effect of churning and blurring in our total sample is on the same order as the effect purely from churning in our circular sample. Thus, beyond the initially large displacements caused by churning in the first hundreds of Myrs, churning begins to function indistinguishably from blurring, spreading only gradually with time. However, this is not to say that the distances covered by churning are less at later times. In fact, stars may be churned back and forth across the spiral patterns co-rotation radius many times in their lifetimes on so-called horseshoe orbits \citep{2002MNRAS.336..785S}. But this movement back and forth contributes little to any net migration when considering the average of the population \citep{2018A&A...616A..86H}, and serves only to broaden the distribution of present-day radii in a diffusive pattern reminiscent of blurring. Our results point to a net migration for a very large fraction of stars.

Previous attempts to constrain the strength of churning using observations from large spectroscopic surveys include \citet{2015MNRAS.447.3526K,2018A&A...609A..79H,2019arXiv190107565H}. These studies have mainly identified potentially migrated stars via the eccentricities of their orbits. \citet{2015MNRAS.447.3526K} found that for stars with super-solar metallicities  about half of the stars in their RAVE sample had $ecc.<0.15$ (which they took to imply a circular orbit). That such stars are present at the solar radius is interpreted as evidence for churning as the ISM is at solar metallicity today and hence, more metal-rich stars need to come from somewhere else than the solar neighbourhood. \citet{2018A&A...609A..79H} used stars from the \textit{Gaia}-ESO Survey. They selected stars with [Fe/H]$>$0.1\,dex and found that about 20\% of the stars in their sample do not reach the Galacto-centric radius at which they likely formed and can thus have been churned. \citet{2019arXiv190107565H} on the other hand, using a similar approach, find that for stars in the solar neighbourhood observed with GALAH and with $ecc.<0.2$ as many as 70\% has reached that location thanks to churning/radial migration. Compared with our more modest total conservative estimate of about 15\% for our full sample this seems a very large number. On the other hand, \citet{2019arXiv190107565H} includes both high- and low-$\alpha$ stars, whilst we \citep[in agreement with][]{2018ApJ...865...96F} argue that any attempts to constrain churning for $\alpha$-enhanced stars is likely to fail due to the complex nature and formation channel of that stellar component which includes both accretion and turbulent formation scenarios in the early Universe \citep[some examples include][]{2009ApJ...707L...1B,Agertz09b}. 
We note that our definition of a circular orbit is much more stringent than used in these studies. Referring to Fig.\,\ref{fig:sample3} we can see that by instead using a cut in $ecc.$ we would indeed have many more stars, but importantly, say if we cut at 0.15 then that would essentially include a very large portion of the stars with 0.95. We note that if we relax the criterion for circularity to 0.95 (from 0.99) then the difference between all stars and stars on circular orbits in Fig.\,\ref{fig:md} disappears. Meaning that then there is no difference between the two samples in terms of displacement, indeed it is not possible to distinguish between churning and blurring (see also discussion in Sect.\,\ref{sect:anal}). Hence, we conclude that although $ecc.$ might appear as an easily understood measure of the orbital shape the more robust $L_{\rm z}/L_{\rm c}$ is a better indicator of the orbital shape. 

At this point in the discussion it is important to re-iterate two things: 1) our study is of an exploratory nature, we wish to see if we can find means to constrain churning and blurring the stellar populations in the cold stellar disk in the Milky Way, 2) the sample we have used is far from perfect for the purposes. We believe that we have succeeded in showing that there are ways to effectively constrain the size and strength of churning  using simple means to estimate the migratory distances for stars in the cold stellar disk. In this work we have made no attempt to account for the selection effects in our sample. Could some of our conclusions be influenced by this? Our main aim with this work is to establish ways to constrain the strength of churning and blurring. The results depends both on the models used (the radial metallicity gradients for the ISM) as well as the quality of stellar sample, including its physical distribution in the Galaxy. An inherent assumption is that there are no azimuthal changes in the stellar population in the Milky Way. This is likely incorrect, but current data does not allow to address this question. Future data releases from \textit{Gaia} coupled with large spectroscopic surveys as well as dedicated follow-up of, e.g., Cepheids and A and B stars across the Milky Way disk will elevate this problem. 

Figure\,\ref{fig:sample2} f) and g) show that the stellar sample we use mainly is situated just outside the solar circle. Although there are stars between 5 -- 11\,kpc there is a concentration around 9\,kpc. There is also a slight trend between [Fe/H] and present day Galacto-centric radius such that more metal-poor stars are found further out in the Galaxy. We know that the Milky Way stellar disk is well populated also inside the solar circle but we know less about the metallicities of those stars. Taking the data at face-value, we can thus conclude that it is likely that we are missing stars on smaller radii meaning that we will not have such a good view of the radial migration experienced by the stars that are currently inside the solar circle. We think, however, that for the stars beyond the solar circle and inside about 10.5\,kpc we have a pretty good sampling of the stellar population as of today. Thus our inferences about the strength of churning and blurring, as applied to these stars, should hold -- churning is the stronger process and acts early on in the life of the stars. Later churning and blurring both acts as diffusive processes that grow slowly over time. 

Future studies must still look in to how the selection function of the stellar sample influences the results. We have seen that for our sample of cold disk stars drawn from a combination of \textit{Gaia} and APOGEE the results are largely model independent. That might not necessarily be the case with a sample defined in a different way and with a different selection function. 

\section{Conclusions}
\label{sect:conclusion}

In this work we set out to explore the possibility to quantify how many stars have radially migrated in the stellar disk and, eventually, be able to put numbers on the strength and importance of the processes involved. We have taken some first steps on this path by utilising a sample of red giant stars from APOGEE DR14, \textit{Gaia} parallaxes and proper motions, and stellar ages derived from C and N abundances in the stars. This sample has allowed us to quantify  how large a fraction of the stars in the sample have had their orbits changed from initially circular to non-circular, how many remain on circular orbits and how many are on circular orbits that might have been ``churned''. 

We find that a conservative estimate is that about 10\% of the stars in the sample have been churned. This is in contrast to recent studies that have much higher numbers, however, we note that those studies essentially look at stars with super-solar metallicities whilst we study stars of all metallicities. Furthermore, our definition of a highly circular orbit is deliberately conservative. If we instead select the stars have Galacto-centric radii at formation that lays outside of its apo- as well as peri-centre today, we find that about half of the stars have undergone some combination of churning and blurring. 
We estimate that a robust $5-7$\% of stars in our sample have not had their orbits blurred, nor have they been churned. These stars appear at all ages indicating that an individual star may escape these dynamical processes for quite a long time. 
Our study also provides tentative observational support to the assumption that stars at a given radius in the Galaxy follow a tight age-metallicity relation.

Looking towards the future, there are several on-going and upcoming large spectroscopic surveys that would be able to provide data to further explore the relative importance and strengths of churning and blurring in our Galaxy \citep[e.g., WEAVE, 4MOST][and Jin et al. in prep.]{2010SPIE.7735E..7GB,2016SPIE.9908E..1OD,2019Msngr.175....3D}. From \textit{Gaia} we will have the parallaxes and proper motions, which when combined with the radial velocities from the spectroscopic surveys will give the full 6D phase space information needed to calculate stellar orbits. The surveys will also provide the needed metallicity and elemental abundances. Stellar ages are difficult to derive. The best prospects are for turn-off stars \citep{2019MNRAS.482..895S} but to reach large volumes of the Milky Way require us to use red giants. In this study we have made use of stellar ages for red giants derived using elemental abundances of C and N and combinations thereof \citep{2016MNRAS.456.3655M}. It is essential that these and similar methods are further evaluated, developed and validated such that they may be used, at least in a statistical sense, in studies like the one presented here. 

\section*{Acknowledgements}

S.F. was supported by the project grant ''The New Milky Way'' from the Knut and Alice Wallenberg foundation and by the grant 2016-03412 from the Swedish Research Council. O.A. acknowledges support from the Swedish Research Council grant 2014-5791 and the Knut and Alice Wallenberg Foundation.

This work has made use of data from the European Space Agency (ESA) mission
{\it Gaia} (\url{https://www.cosmos.esa.int/gaia}), processed by the {\it Gaia}
Data Processing and Analysis Consortium (DPAC,
\url{https://www.cosmos.esa.int/web/gaia/dpac/consortium}). Funding for the DPAC
has been provided by national institutions, in particular the institutions
participating in the {\it Gaia} Multilateral Agreement.

Funding for the Sloan Digital Sky Survey IV has been provided by the Alfred P. Sloan Foundation, the U.S. Department of Energy Office of Science, and the Participating Institutions. SDSS-IV acknowledges
support and resources from the Center for High-Performance Computing at
the University of Utah. The SDSS web site is www.sdss.org.

SDSS-IV is managed by the Astrophysical Research Consortium for the 
Participating Institutions of the SDSS Collaboration including the 
Brazilian Participation Group, the Carnegie Institution for Science, 
Carnegie Mellon University, the Chilean Participation Group, the French Participation Group, Harvard-Smithsonian Center for Astrophysics, 
Instituto de Astrof\'isica de Canarias, The Johns Hopkins University, Kavli Institute for the Physics and Mathematics of the Universe (IPMU) / 
University of Tokyo, the Korean Participation Group, Lawrence Berkeley National Laboratory, 
Leibniz Institut f\"ur Astrophysik Potsdam (AIP),  
Max-Planck-Institut f\"ur Astronomie (MPIA Heidelberg), 
Max-Planck-Institut f\"ur Astrophysik (MPA Garching), 
Max-Planck-Institut f\"ur Extraterrestrische Physik (MPE), 
National Astronomical Observatories of China, New Mexico State University, 
New York University, University of Notre Dame, 
Observat\'ario Nacional / MCTI, The Ohio State University, 
Pennsylvania State University, Shanghai Astronomical Observatory, 
United Kingdom Participation Group,
Universidad Nacional Aut\'onoma de M\'exico, University of Arizona, 
University of Colorado Boulder, University of Oxford, University of Portsmouth, 
University of Utah, University of Virginia, University of Washington, University of Wisconsin, 
Vanderbilt University, and Yale University.



\bibliographystyle{mnras}
\bibliography{references} 

\begin{thebibliography}{}
\makeatletter
\relax
\def\mn@urlcharsother{\let\do\@makeother \do\$\do\&\do\#\do\^\do\_\do\%\do\~}
\def\mn@doi{\begingroup\mn@urlcharsother \@ifnextchar [ {\mn@doi@}
  {\mn@doi@[]}}
\def\mn@doi@[#1]#2{\def\@tempa{#1}\ifx\@tempa\@empty \href
  {http://dx.doi.org/#2} {doi:#2}\else \href {http://dx.doi.org/#2} {#1}\fi
  \endgroup}
\def\mn@eprint#1#2{\mn@eprint@#1:#2::\@nil}
\def\mn@eprint@arXiv#1{\href {http://arxiv.org/abs/#1} {{\tt arXiv:#1}}}
\def\mn@eprint@dblp#1{\href {http://dblp.uni-trier.de/rec/bibtex/#1.xml}
  {dblp:#1}}
\def\mn@eprint@#1:#2:#3:#4\@nil{\def\@tempa {#1}\def\@tempb {#2}\def\@tempc
  {#3}\ifx \@tempc \@empty \let \@tempc \@tempb \let \@tempb \@tempa \fi \ifx
  \@tempb \@empty \def\@tempb {arXiv}\fi \@ifundefined
  {mn@eprint@\@tempb}{\@tempb:\@tempc}{\expandafter \expandafter \csname
  mn@eprint@\@tempb\endcsname \expandafter{\@tempc}}}

\bibitem[\protect\citeauthoryear{{Agertz}, {Teyssier}  \& {Moore}}{{Agertz}
  et~al.}{2009}]{Agertz09b}
{Agertz} O.,  {Teyssier} R.,   {Moore} B.,  2009, \mn@doi [\mnras]
  {10.1111/j.1745-3933.2009.00685.x}, \href
  {http://adsabs.harvard.edu/abs/2009MNRAS.397L..64A} {397, L64}

\bibitem[\protect\citeauthoryear{{Alam} et~al.,}{{Alam}
  et~al.}{2015}]{2015ApJS..219...12A}
{Alam} S.,  et~al., 2015, \mn@doi [\apjs] {10.1088/0067-0049/219/1/12}, \href
  {https://ui.adsabs.harvard.edu/abs/2015ApJS..219...12A} {219, 12}

\bibitem[\protect\citeauthoryear{{Balcells} et~al.,}{{Balcells}
  et~al.}{2010}]{2010SPIE.7735E..7GB}
{Balcells} M.,  et~al., 2010, in Ground-based and Airborne Instrumentation for
  Astronomy III. p. 77357G (\mn@eprint {arXiv} {1008.0600}),
  \mn@doi{10.1117/12.856947}

\bibitem[\protect\citeauthoryear{{Bensby}, {Feltzing}  \& {Oey}}{{Bensby}
  et~al.}{2014}]{2014A&A...562A..71B}
{Bensby} T.,  {Feltzing} S.,   {Oey} M.~S.,  2014, \mn@doi [\aap]
  {10.1051/0004-6361/201322631}, \href
  {http://adsabs.harvard.edu/abs/2014A%26A...562A..71B} {562, A71}

\bibitem[\protect\citeauthoryear{{Bissantz}, {Englmaier}  \&
  {Gerhard}}{{Bissantz} et~al.}{2003}]{2003MNRAS.340..949B}
{Bissantz} N.,  {Englmaier} P.,   {Gerhard} O.,  2003, \mn@doi [\mnras]
  {10.1046/j.1365-8711.2003.06358.x}, \href
  {https://ui.adsabs.harvard.edu/abs/2003MNRAS.340..949B} {340, 949}

\bibitem[\protect\citeauthoryear{{Bournaud}, {Elmegreen}  \&
  {Martig}}{{Bournaud} et~al.}{2009}]{2009ApJ...707L...1B}
{Bournaud} F.,  {Elmegreen} B.~G.,   {Martig} M.,  2009, \mn@doi [\apjl]
  {10.1088/0004-637X/707/1/L1}, \href
  {https://ui.adsabs.harvard.edu/abs/2009ApJ...707L...1B} {707, L1}

\bibitem[\protect\citeauthoryear{{Bovy}}{{Bovy}}{2015}]{2015ApJS..216...29B}
{Bovy} J.,  2015, \mn@doi [\apjs] {10.1088/0067-0049/216/2/29}, \href
  {http://adsabs.harvard.edu/abs/2015ApJS..216...29B} {216, 29}

\bibitem[\protect\citeauthoryear{{Casagrande}, {Sch{\"o}nrich}, {Asplund},
  {Cassisi}, {Ram{\'\i}rez}, {Mel{\'e}ndez}, {Bensby}  \&
  {Feltzing}}{{Casagrande} et~al.}{2011}]{2011A&A...530A.138C}
{Casagrande} L.,  {Sch{\"o}nrich} R.,  {Asplund} M.,  {Cassisi} S.,
  {Ram{\'\i}rez} I.,  {Mel{\'e}ndez} J.,  {Bensby} T.,   {Feltzing} S.,  2011,
  \mn@doi [\aap] {10.1051/0004-6361/201016276}, \href
  {https://ui.adsabs.harvard.edu/abs/2011A&A...530A.138C} {530, A138}

\bibitem[\protect\citeauthoryear{{Chakrabarty}}{{Chakrabarty}}{2007}]{2007A&A...467..145C}
{Chakrabarty} D.,  2007, \mn@doi [\aap] {10.1051/0004-6361:20066677}, \href
  {https://ui.adsabs.harvard.edu/abs/2007A&A...467..145C} {467, 145}

\bibitem[\protect\citeauthoryear{{Dehnen}}{{Dehnen}}{2000}]{2000AJ....119..800D}
{Dehnen} W.,  2000, \mn@doi [\aj] {10.1086/301226}, \href
  {http://adsabs.harvard.edu/abs/2000AJ....119..800D} {119, 800}

\bibitem[\protect\citeauthoryear{{Edvardsson}, {Andersen}, {Gustafsson},
  {Lambert}, {Nissen}  \& {Tomkin}}{{Edvardsson}
  et~al.}{1993}]{1993A&A...275..101E}
{Edvardsson} B.,  {Andersen} J.,  {Gustafsson} B.,  {Lambert} D.~L.,  {Nissen}
  P.~E.,   {Tomkin} J.,  1993, \aap, \href
  {https://ui.adsabs.harvard.edu/abs/1993A&A...275..101E} {500, 391}

\bibitem[\protect\citeauthoryear{{Feltzing}, {Holmberg}  \&
  {Hurley}}{{Feltzing} et~al.}{2001}]{2001A&A...377..911F}
{Feltzing} S.,  {Holmberg} J.,   {Hurley} J.~R.,  2001, \mn@doi [\aap]
  {10.1051/0004-6361:20011119}, \href
  {https://ui.adsabs.harvard.edu/abs/2001A&A...377..911F} {377, 911}

\bibitem[\protect\citeauthoryear{{Fern{\'a}ndez}, {Figueras}  \&
  {Torra}}{{Fern{\'a}ndez} et~al.}{2001}]{2001A&A...372..833F}
{Fern{\'a}ndez} D.,  {Figueras} F.,   {Torra} J.,  2001, \mn@doi [\aap]
  {10.1051/0004-6361:20010366}, \href
  {https://ui.adsabs.harvard.edu/abs/2001A&A...372..833F} {372, 833}

\bibitem[\protect\citeauthoryear{{Frankel}, {Rix}, {Ting}, {Ness}  \&
  {Hogg}}{{Frankel} et~al.}{2018}]{2018ApJ...865...96F}
{Frankel} N.,  {Rix} H.-W.,  {Ting} Y.-S.,  {Ness} M.,   {Hogg} D.~W.,  2018,
  \mn@doi [\apj] {10.3847/1538-4357/aadba5}, \href
  {http://adsabs.harvard.edu/abs/2018ApJ...865...96F} {865, 96}

\bibitem[\protect\citeauthoryear{{Fuhrmann}, {Chini}, {Kaderhandt}  \&
  {Chen}}{{Fuhrmann} et~al.}{2017}]{2017MNRAS.464.2610F}
{Fuhrmann} K.,  {Chini} R.,  {Kaderhandt} L.,   {Chen} Z.,  2017, \mn@doi
  [\mnras] {10.1093/mnras/stw2526}, \href
  {https://ui.adsabs.harvard.edu/abs/2017MNRAS.464.2610F} {464, 2610}

\bibitem[\protect\citeauthoryear{{Gaia Collaboration} et~al.,}{{Gaia
  Collaboration} et~al.}{2016}]{2016A&A...595A...1G}
{Gaia Collaboration} et~al., 2016, \mn@doi [\aap]
  {10.1051/0004-6361/201629272}, \href
  {http://adsabs.harvard.edu/abs/2016A%26A...595A...1G} {595, A1}

\bibitem[\protect\citeauthoryear{{Gaia Collaboration} et~al.,}{{Gaia
  Collaboration} et~al.}{2018a}]{2018A&A...616A...1G}
{Gaia Collaboration} et~al., 2018a, \mn@doi [\aap]
  {10.1051/0004-6361/201833051}, \href
  {http://adsabs.harvard.edu/abs/2018A%26A...616A...1G} {616, A1}

\bibitem[\protect\citeauthoryear{{Gaia Collaboration} et~al.,}{{Gaia
  Collaboration} et~al.}{2018b}]{2018A&A...616A..10G}
{Gaia Collaboration} et~al., 2018b, \mn@doi [\aap]
  {10.1051/0004-6361/201832843}, \href
  {https://ui.adsabs.harvard.edu/abs/2018A&A...616A..10G} {616, A10}

\bibitem[\protect\citeauthoryear{{Gerhard}}{{Gerhard}}{2011}]{2011MSAIS..18..185G}
{Gerhard} O.,  2011, Memorie della Societa Astronomica Italiana Supplementi,
  \href {https://ui.adsabs.harvard.edu/abs/2011MSAIS..18..185G} {18, 185}

\bibitem[\protect\citeauthoryear{{Grand} \& {Kawata}}{{Grand} \&
  {Kawata}}{2016}]{2016AN....337..957G}
{Grand} R.~J.~J.,  {Kawata} D.,  2016, \mn@doi [Astronomische Nachrichten]
  {10.1002/asna.201612407}, \href
  {http://adsabs.harvard.edu/abs/2016AN....337..957G} {337, 957}

\bibitem[\protect\citeauthoryear{{Grenon}}{{Grenon}}{1987}]{1987JApA....8..123G}
{Grenon} M.,  1987, \mn@doi [Journal of Astrophysics and Astronomy]
  {10.1007/BF02714310}, \href
  {https://ui.adsabs.harvard.edu/abs/1987JApA....8..123G} {8, 123}

\bibitem[\protect\citeauthoryear{{Halle}, {Di Matteo}, {Haywood}  \&
  {Combes}}{{Halle} et~al.}{2018}]{2018A&A...616A..86H}
{Halle} A.,  {Di Matteo} P.,  {Haywood} M.,   {Combes} F.,  2018, \mn@doi
  [\aap] {10.1051/0004-6361/201832603}, \href
  {http://adsabs.harvard.edu/abs/2018A%26A...616A..86H} {616, A86}

\bibitem[\protect\citeauthoryear{{Hayden} et~al.,}{{Hayden}
  et~al.}{2018}]{2018A&A...609A..79H}
{Hayden} M.~R.,  et~al., 2018, \mn@doi [\aap] {10.1051/0004-6361/201730412},
  \href {https://ui.adsabs.harvard.edu/abs/2018A&A...609A..79H} {609, A79}

\bibitem[\protect\citeauthoryear{{Hayden} et~al.,}{{Hayden}
  et~al.}{2019}]{2019arXiv190107565H}
{Hayden} M.~R.,  et~al., 2019, arXiv e-prints, \href
  {https://ui.adsabs.harvard.edu/abs/2019arXiv190107565H} {p. arXiv:1901.07565}

\bibitem[\protect\citeauthoryear{{Haywood}, {Snaith}, {Lehnert}, {Di Matteo}
  \& {Khoperskov}}{{Haywood} et~al.}{2019}]{2019A&A...625A.105H}
{Haywood} M.,  {Snaith} O.,  {Lehnert} M.~D.,  {Di Matteo} P.,   {Khoperskov}
  S.,  2019, \mn@doi [\aap] {10.1051/0004-6361/201834155}, \href
  {https://ui.adsabs.harvard.edu/abs/2019A&A...625A.105H} {625, A105}

\bibitem[\protect\citeauthoryear{{Kassin} et~al.,}{{Kassin}
  et~al.}{2012}]{2012ApJ...758..106K}
{Kassin} S.~A.,  et~al., 2012, \mn@doi [\apj] {10.1088/0004-637X/758/2/106},
  \href {https://ui.adsabs.harvard.edu/abs/2012ApJ...758..106K} {758, 106}

\bibitem[\protect\citeauthoryear{{Kordopatis} et~al.,}{{Kordopatis}
  et~al.}{2015}]{2015MNRAS.447.3526K}
{Kordopatis} G.,  et~al., 2015, \mn@doi [\mnras] {10.1093/mnras/stu2726}, \href
  {https://ui.adsabs.harvard.edu/abs/2015MNRAS.447.3526K} {447, 3526}

\bibitem[\protect\citeauthoryear{{Kubryk}, {Prantzos}  \&
  {Athanassoula}}{{Kubryk} et~al.}{2013}]{2013MNRAS.436.1479K}
{Kubryk} M.,  {Prantzos} N.,   {Athanassoula} E.,  2013, \mn@doi [\mnras]
  {10.1093/mnras/stt1667}, \href
  {https://ui.adsabs.harvard.edu/abs/2013MNRAS.436.1479K} {436, 1479}

\bibitem[\protect\citeauthoryear{{Kubryk}, {Prantzos}  \&
  {Athanassoula}}{{Kubryk} et~al.}{2015a}]{2015AA...580A.126K}
{Kubryk} M.,  {Prantzos} N.,   {Athanassoula} E.,  2015a, \mn@doi [\aap]
  {10.1051/0004-6361/201424171}, \href
  {http://adsabs.harvard.edu/abs/2015A%26A...580A.126K} {580, A126}

\bibitem[\protect\citeauthoryear{{Kubryk}, {Prantzos}  \&
  {Athanassoula}}{{Kubryk} et~al.}{2015b}]{2015A&A...580A.127K}
{Kubryk} M.,  {Prantzos} N.,   {Athanassoula} E.,  2015b, \mn@doi [\aap]
  {10.1051/0004-6361/201424599}, \href
  {http://adsabs.harvard.edu/abs/2015A%26A...580A.127K} {580, A127}

\bibitem[\protect\citeauthoryear{{Liu} \& {van de Ven}}{{Liu} \& {van de
  Ven}}{2012}]{2012MNRAS.425.2144L}
{Liu} C.,  {van de Ven} G.,  2012, \mn@doi [\mnras]
  {10.1111/j.1365-2966.2012.21551.x}, \href
  {http://adsabs.harvard.edu/abs/2012MNRAS.425.2144L} {425, 2144}

\bibitem[\protect\citeauthoryear{{Loebman}, {Ro{\v s}kar}, {Debattista},
  {Ivezi{\'c}}, {Quinn}  \& {Wadsley}}{{Loebman}
  et~al.}{2011}]{2011ApJ...737....8L}
{Loebman} S.~R.,  {Ro{\v s}kar} R.,  {Debattista} V.~P.,  {Ivezi{\'c}} {\v Z}.,
   {Quinn} T.~R.,   {Wadsley} J.,  2011, \mn@doi [\apj]
  {10.1088/0004-637X/737/1/8}, \href
  {https://ui.adsabs.harvard.edu/abs/2011ApJ...737....8L} {737, 8}

\bibitem[\protect\citeauthoryear{{Majewski} et~al.,}{{Majewski}
  et~al.}{2017}]{2017AJ....154...94M}
{Majewski} S.~R.,  et~al., 2017, \mn@doi [\aj] {10.3847/1538-3881/aa784d},
  \href {http://adsabs.harvard.edu/abs/2017AJ....154...94M} {154, 94}

\bibitem[\protect\citeauthoryear{{Martig} et~al.,}{{Martig}
  et~al.}{2016}]{2016MNRAS.456.3655M}
{Martig} M.,  et~al., 2016, \mn@doi [\mnras] {10.1093/mnras/stv2830}, \href
  {https://ui.adsabs.harvard.edu/abs/2016MNRAS.456.3655M} {456, 3655}

\bibitem[\protect\citeauthoryear{{Mart{\'\i}nez-Barbosa}, {Brown}, {Boekholt},
  {Portegies Zwart}, {Antiche}  \& {Antoja}}{{Mart{\'\i}nez-Barbosa}
  et~al.}{2016}]{2016MNRAS.457.1062M}
{Mart{\'\i}nez-Barbosa} C.~A.,  {Brown} A.~G.~A.,  {Boekholt} T.,  {Portegies
  Zwart} S.,  {Antiche} E.,   {Antoja} T.,  2016, \mn@doi [\mnras]
  {10.1093/mnras/stw006}, \href
  {https://ui.adsabs.harvard.edu/abs/2016MNRAS.457.1062M} {457, 1062}

\bibitem[\protect\citeauthoryear{{Minchev}, {Chiappini}  \& {Martig}}{{Minchev}
  et~al.}{2013}]{2013A&A...558A...9M}
{Minchev} I.,  {Chiappini} C.,   {Martig} M.,  2013, \mn@doi [\aap]
  {10.1051/0004-6361/201220189}, \href
  {https://ui.adsabs.harvard.edu/abs/2013A&A...558A...9M} {558, A9}

\bibitem[\protect\citeauthoryear{{Minchev} et~al.,}{{Minchev}
  et~al.}{2018}]{2018MNRAS.481.1645M}
{Minchev} I.,  et~al., 2018, \mn@doi [\mnras] {10.1093/mnras/sty2033}, \href
  {http://adsabs.harvard.edu/abs/2018MNRAS.481.1645M} {481, 1645}

\bibitem[\protect\citeauthoryear{{Pehlivan Rhodin}, {Agertz}, {Christensen},
  {Renaud}  \& {Uldall Fynbo}}{{Pehlivan Rhodin} et~al.}{2019}]{Rhodin2019}
{Pehlivan Rhodin} N.~H.,  {Agertz} O.,  {Christensen} L.,  {Renaud} F.,
  {Uldall Fynbo} J.~P.,  2019, arXiv e-prints, \href
  {https://ui.adsabs.harvard.edu/abs/2019arXiv190110777P} {}

\bibitem[\protect\citeauthoryear{{Ro{\v s}kar}, {Debattista}, {Quinn},
  {Stinson}  \& {Wadsley}}{{Ro{\v s}kar} et~al.}{2008}]{2008ApJ...684L..79R}
{Ro{\v s}kar} R.,  {Debattista} V.~P.,  {Quinn} T.~R.,  {Stinson} G.~S.,
  {Wadsley} J.,  2008, \mn@doi [\apjl] {10.1086/592231}, \href
  {http://adsabs.harvard.edu/abs/2008ApJ...684L..79R} {684, L79}

\bibitem[\protect\citeauthoryear{{Sahlholdt}, {Feltzing}, {Lindegren}  \&
  {Church}}{{Sahlholdt} et~al.}{2019}]{2019MNRAS.482..895S}
{Sahlholdt} C.~L.,  {Feltzing} S.,  {Lindegren} L.,   {Church} R.~P.,  2019,
  \mn@doi [\mnras] {10.1093/mnras/sty2732}, \href
  {https://ui.adsabs.harvard.edu/abs/2019MNRAS.482..895S} {482, 895}

\bibitem[\protect\citeauthoryear{{Sanders} \& {Binney}}{{Sanders} \&
  {Binney}}{2015}]{2015MNRAS.449.3479S}
{Sanders} J.~L.,  {Binney} J.,  2015, \mn@doi [\mnras] {10.1093/mnras/stv578},
  \href {http://adsabs.harvard.edu/abs/2015MNRAS.449.3479S} {449, 3479}

\bibitem[\protect\citeauthoryear{{Sch{\"o}nrich} \& {Binney}}{{Sch{\"o}nrich}
  \& {Binney}}{2009}]{2009MNRAS.396..203S}
{Sch{\"o}nrich} R.,  {Binney} J.,  2009, \mn@doi [\mnras]
  {10.1111/j.1365-2966.2009.14750.x}, \href
  {http://adsabs.harvard.edu/abs/2009MNRAS.396..203S} {396, 203}

\bibitem[\protect\citeauthoryear{{Sellwood} \& {Binney}}{{Sellwood} \&
  {Binney}}{2002}]{2002MNRAS.336..785S}
{Sellwood} J.~A.,  {Binney} J.~J.,  2002, \mn@doi [\mnras]
  {10.1046/j.1365-8711.2002.05806.x}, \href
  {http://adsabs.harvard.edu/abs/2002MNRAS.336..785S} {336, 785}

\bibitem[\protect\citeauthoryear{{Teyssier}}{{Teyssier}}{2002}]{teyssier02}
{Teyssier} R.,  2002, \mn@doi [\aap] {10.1051/0004-6361:20011817}, \href
  {http://adsabs.harvard.edu/abs/2002A%26A...385..337T} {385, 337}

\bibitem[\protect\citeauthoryear{{Wielen}}{{Wielen}}{1977}]{1977A&A....60..263W}
{Wielen} R.,  1977, \aap, \href
  {https://ui.adsabs.harvard.edu/abs/1977A&A....60..263W} {60, 263}

\bibitem[\protect\citeauthoryear{{Wielen}, {Fuchs}  \& {Dettbarn}}{{Wielen}
  et~al.}{1996}]{1996A&A...314..438W}
{Wielen} R.,  {Fuchs} B.,   {Dettbarn} C.,  1996, \aap, \href
  {https://ui.adsabs.harvard.edu/abs/1996A&A...314..438W} {314, 438}

\bibitem[\protect\citeauthoryear{{de Jong} et~al.,}{{de Jong}
  et~al.}{2016}]{2016SPIE.9908E..1OD}
{de Jong} R.~S.,  et~al., 2016, in Ground-based and Airborne Instrumentation
  for Astronomy VI. p. 99081O, \mn@doi{10.1117/12.2232832}

\bibitem[\protect\citeauthoryear{{de Jong} et~al.,}{{de Jong}
  et~al.}{2019}]{2019Msngr.175....3D}
{de Jong} R.~S.,  et~al., 2019, \mn@doi [The Messenger]
  {10.18727/0722-6691/5117}, \href
  {https://ui.adsabs.harvard.edu/abs/2019Msngr.175....3D} {175, 3}

\makeatother
\end{thebibliography}


\appendix

\section{Radial metallicity gradients for the ISM -- equations}

\subsection{\citet{2018ApJ...865...96F} radial metallicity gradients for the ISM}
\label{app:iso2}

\begin{figure}
	\includegraphics[width=\columnwidth]{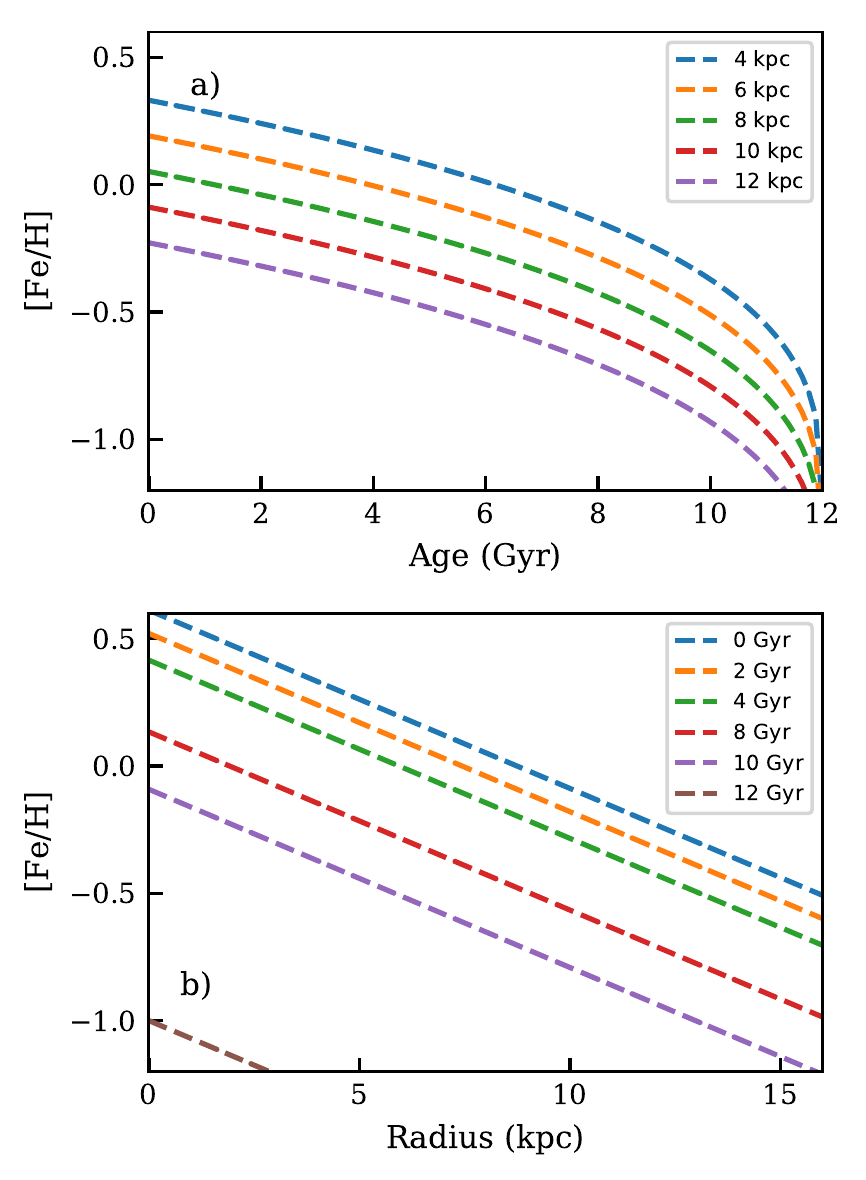}
\caption{{\bf a)} [Fe/H] as a function of age for different Galactocentric distances from \citet{2018ApJ...865...96F} (compare their Fig X). {\bf b)} [Fe/H] as a function of Galactocentric radius for different ages, Eq.\,(\ref{eq:frankel}).}
    \label{appfig:iso2}
\end{figure}

Figure\,\ref{appfig:iso2} {\bf a} shows the relation of [Fe/H] as a function of age for different Galactocentric radii based on the prescriptions given in \citet{2018ApJ...865...96F}. 

Manipulating the equations in \citet{2018ApJ...865...96F} we arrive at the following equation that describes how the iron content in the ISM changes with radius ($R$) for a given age ($\tau$):

\begin{equation}
[Fe/H] (\tau) = -1 - (-1 -0.07\cdot 8.74)\cdot ((1 - \tau /12)^{0.32}) -0.07 \cdot R
\label{eq:frankel}
\end{equation}

Figure\,\ref{appfig:iso2} {\bf b} then shows the resulting radial gradients used in our work. For full references and discussion of the constants in Eq.(\ref{eq:frankel}) we refer the reader to \citet{2018ApJ...865...96F}.

\subsection{\citet{2015MNRAS.449.3479S} radial metallicity gradients for the ISM}
\label{app:iso3}

\begin{figure}
	\includegraphics[width=\columnwidth]{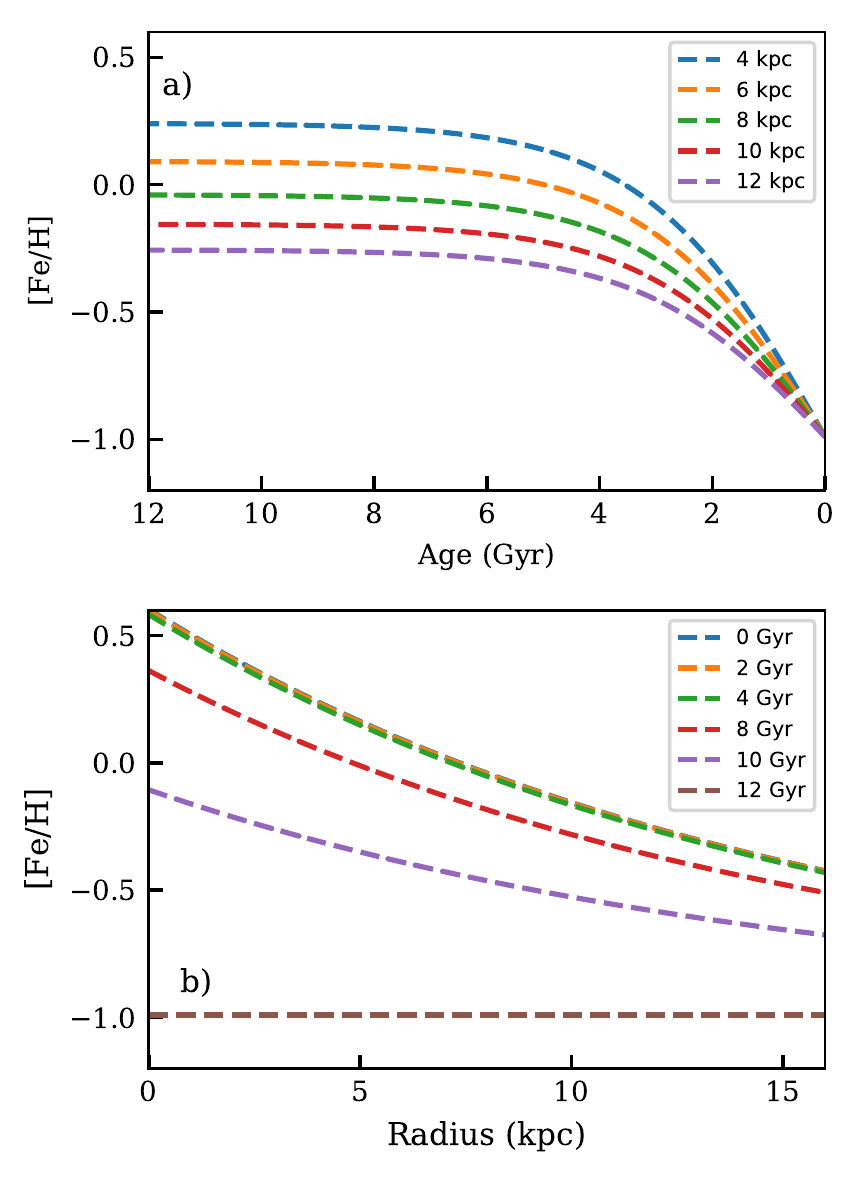}
\caption{{\bf a)} [Fe/H] as a function of age for different Galacto-centric distances from \citet{2015MNRAS.449.3479S} (compare their Fig.\,1). Note that the $x$-axis labelling is reversed as compared to the same plot in Fig.\,\ref{appfig:iso2}. Here 12\,Gyr is "now". {\bf b)} [Fe/H] as a function of Galacto-centric radius for different ages, Eq.\,(\ref{eq:s2}).}
    \label{appfig:iso3}
\end{figure}

Figure\,\ref{appfig:iso3} {\bf a} shows the relation of [Fe/H] as a function of age for different Galacto-centric radii based on the prescriptions given in \citet{2015MNRAS.449.3479S} (their Figure\,1). \citet{2015MNRAS.449.3479S} obtain their functional form for $F(R)$ by fitting to the resulting output from \citep{2009MNRAS.396..203S}, which has a steep present day gradient. $-0.082$\,dex\,kpc$^{-1}$ as compared to $-0.07$\,dex\,kpc$^{-1}$ in most other models \citep[e.g.,][]{2018MNRAS.481.1645M,2018ApJ...865...96F}.

Manipulating the equations in \citet{2015MNRAS.449.3479S} we arrive at the following equation that describes how the iron content in the ISM changes with radius ($R$) for a given age ($\tau$):

\begin{equation}
F(R) = -0.99 \cdot (1- exp(-0.064 \cdot (R-7.37)/0.99))
\label{eq:s1}
\end{equation}

\begin{equation}
[Fe/H] (\tau) = (F(R) + 0.99)\cdot tanh((12- \tau)/3.2) - 0.99
\label{eq:s2}
\end{equation}

Figure\,\ref{appfig:iso3} {\bf b} then shows the resulting radial gradients used in our work. For full references and discussion of the constants in Eq.(\ref{eq:s1} and \ref{eq:s2}) we refer the reader to \citet{2015MNRAS.449.3479S}. We note that as opposed to the other models used in our study this one hardly has any evolution of the ISM radial gradient with time for the time span that is of interest for the formation of the stars in the cold stellar disk, i.e. the last 6 to 8 Gyrs.


\section{Orbital data}

As described in Sect.\,\ref{sect:orbits} orbital data for the stars in our sample were calculated using the {\tt galpy}-package \citep{2015ApJS..216...29B}. 

\subsection{Comparing $L_{\rm z}/L_{\rm c}$ and $ecc.$}
\label{sect:lecc}
\begin{figure}
	\includegraphics[width=\columnwidth]{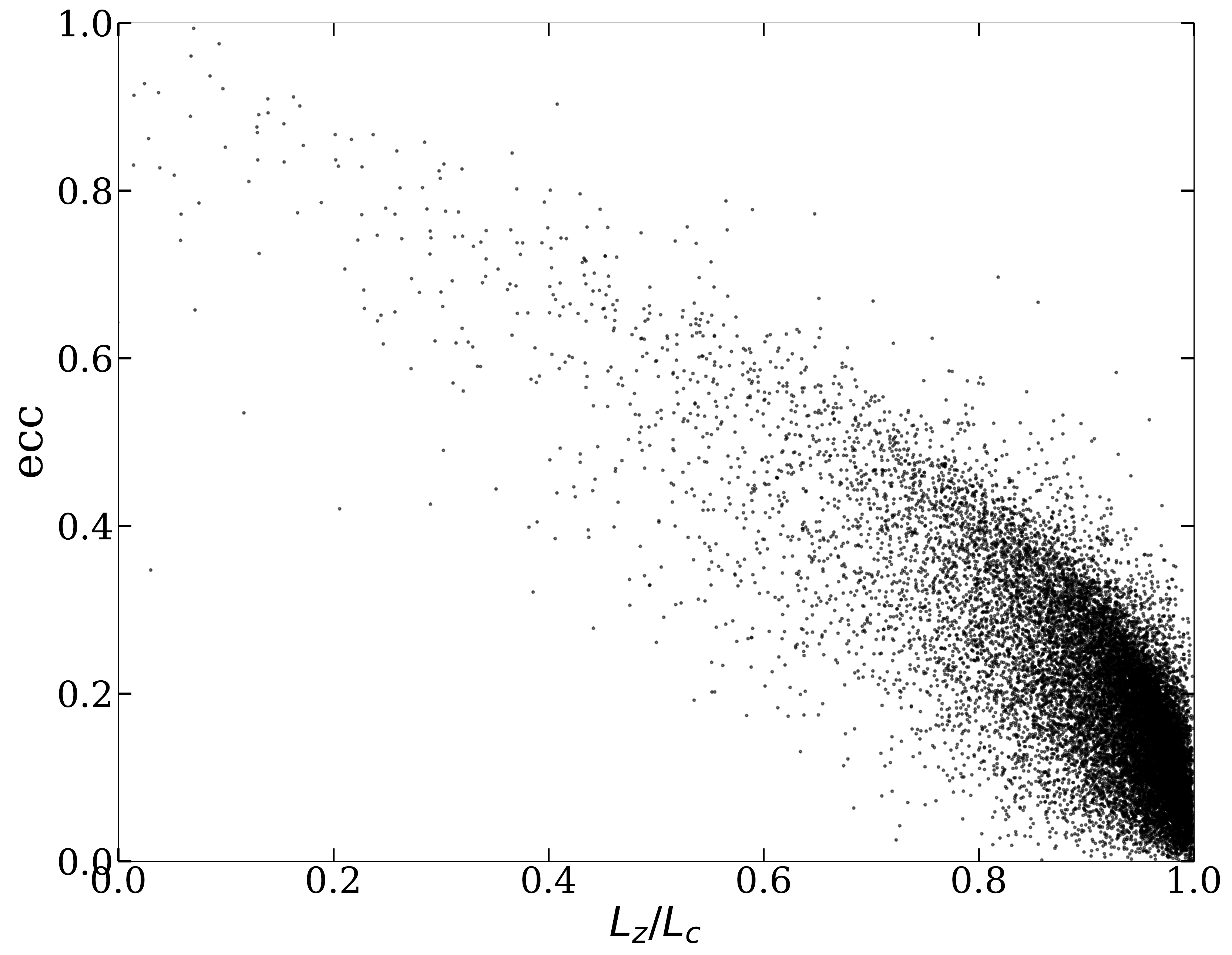}
    \caption{Orbital eccentricity as a function of $L_{\rm z}/L_{\rm c}$ for our sample.}
    \label{fig:sample3}
\end{figure}

In this work, following \citet{2012MNRAS.425.2144L}, we have chosen to use $L_{\rm z}/L_{\rm c}$ to characterise the 
circularity of the stellar orbits.   
Figure\,\ref{fig:sample3} shows a comparison $L_{\rm z}/L_{\rm c}$ and $ecc.$. There is a clear correlation between the two properties. At a given $L_{\rm z}/L_{\rm c}$ there is a substantial spread in $ecc.$. We note that our chosen cut for defining very circular orbits is 0.99 for $L_{\rm z}/L_{\rm c}$, this encompasses $ecc.$ in the range 0 to 0.2.

\subsection{Including a bar in the Galactic potential}
\begin{figure}
	\includegraphics[width=\columnwidth]{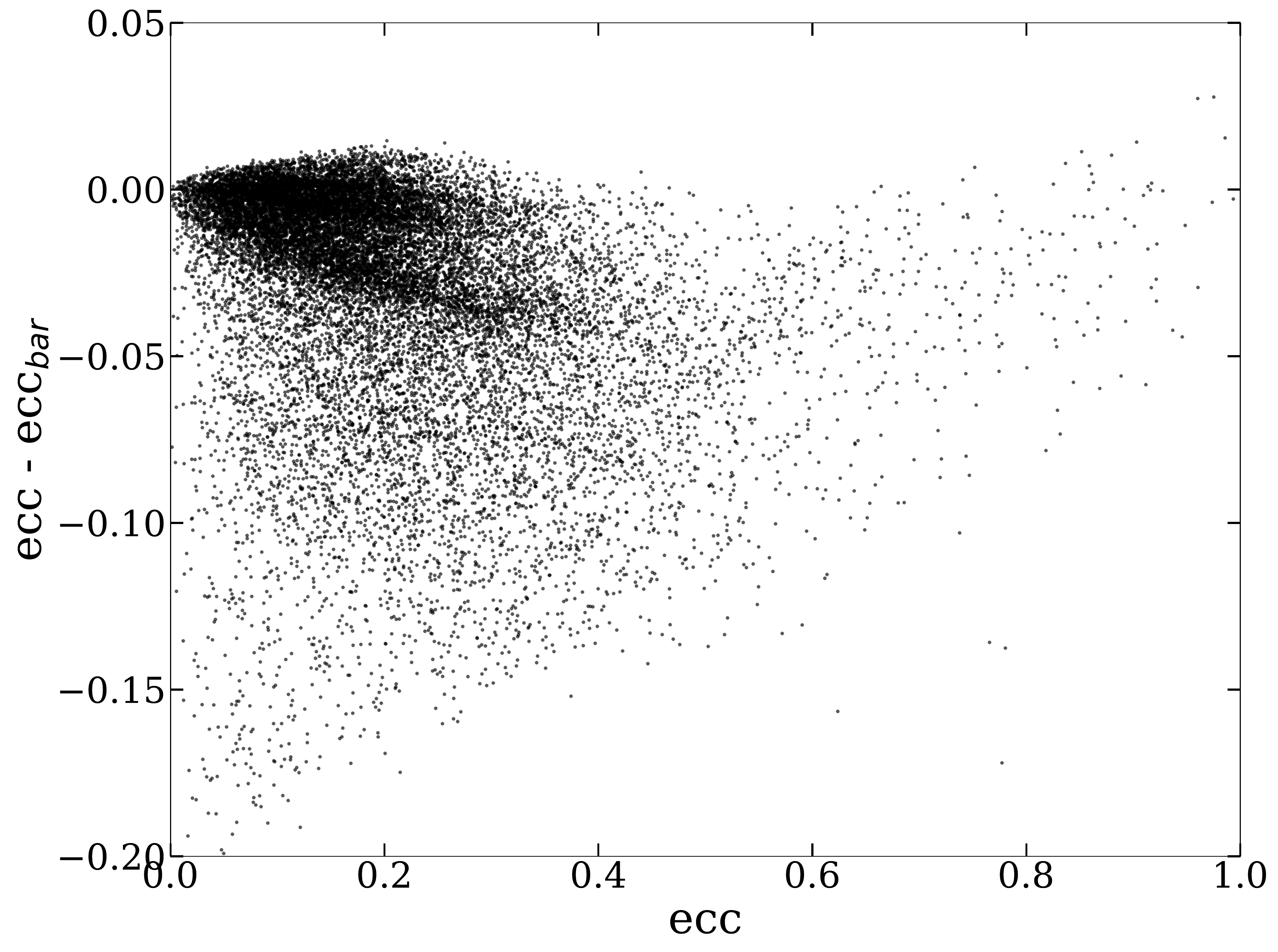}
\caption{Comparison of orbital eccentricities in a potential with and without a bar.}
    \label{appfig:ecc}
\end{figure}

The {\tt galpy}-package allows the user to include a bar when carrying out the orbital integrations \citep[the reader is referred to][for details about the potential]{2015ApJS..216...29B} . Figure\,\ref{appfig:ecc} shows the effect on $ecc.$ when the bar is included in the potential. Although there are noticeable effects we note that for stars that we consider to be on circular orbits (which all have $<0.2$, see Sect.\,\ref{sect:lecc}) the effect is small and hence we have concluded that we did not need to include a bar in the Galactic potential for the purpose of this study.

\subsection{$L_{\rm z}/L_{\rm c} < 0.95$}
\label{app:95}

\begin{table*}
	\caption{Fractions of stars on different types of orbits for different age bins and for 
	all stars (last column). See Sect.\,\ref{sect:churn} for a description. }
	\label{tab:result2}
	\begin{tabular}{lccccc} 
		\hline
	& \multicolumn{5}{l}{Fraction of stars that have $L_{\rm z}/L_{\rm c} > 0.95$ }\\
		Model & Age $<2$ & $2<$ Age $<4$ & $4<$ Age $<6$ & $6<$ Age $<8$ & All ages\\
		\hline
		\citet{2018MNRAS.481.1645M} & 0.747$\pm$0.008 & 0.674$\pm$0.008 & 0.624$\pm$0.011 & 0.560$\pm$0.020 & 0.670$\pm$0.002\\
		\citet{2018ApJ...865...96F} & 0.755$\pm$0.008 & 0.688$\pm$0.008 & 0.649$\pm$0.012 & 0.617$\pm$0.024 & 0.695$\pm$0.003\\
		\citet{2015MNRAS.449.3479S} & 0.763$\pm$0.008 & 0.684$\pm$0.008 & 0.623$\pm$0.012 & 0.569$\pm$0.023 & 0.684$\pm$0.002\\
		\citet{2015AA...580A.126K}  & 0.741$\pm$0.009 & 0.671$\pm$0.007 & 0.627$\pm$0.010 & 0.604$\pm$0.021 & 0.669$\pm$0.003\\
		\hline
	& \multicolumn{5}{l}{Fraction of stars that have $L_{\rm z}/L_{\rm c} > 0.95$  and outside present day orbit}\\
		Model & Age $<2$ & $2<$ Age $<4$ & $4<$ Age $<6$ & $6<$ Age $<8$ & All ages\\
		\hline
		\citet{2018MNRAS.481.1645M} & 0.339$\pm$0.009 & 0.370$\pm$0.008 & 0.385$\pm$0.012 & 0.372$\pm$0.020 & 0.366$\pm$0.004\\
		\citet{2018ApJ...865...96F} & 0.414$\pm$0.009 & 0.442$\pm$0.009 & 0.457$\pm$0.013 & 0.453$\pm$0.025 & 0.453$\pm$0.005\\
		\citet{2015MNRAS.449.3479S} & 0.523$\pm$0.009 & 0.436$\pm$0.009 & 0.386$\pm$0.013 & 0.373$\pm$0.023 & 0.373$\pm$0.003\\
		\citet{2015AA...580A.126K}  & 0.302$\pm$0.010 & 0.332$\pm$0.008 & 0.415$\pm$0.011 & 0.498$\pm$0.022 & 0.498$\pm$0.005\\
	\hline			
	\end{tabular}
\end{table*}

For completeness we include here the resulting fraction of stars on circular orbits when the constraint has been reduced to 0.95 instead of the 0.99 we use in the final analysis (see Table\,\ref{tab:result} for the 0.99 results). The data are given in Table\,\ref{tab:result2} and are also shown in Fig.\,\ref{fig:churn}.


\bsp	
\label{lastpage}
\end{document}